\documentclass[reqno]{amsart}

\usepackage[english]{babel}
\usepackage[a4paper, total={7in, 9.5in}]{geometry}

\usepackage{amsmath}

\usepackage{chemformula}
\usepackage{graphicx}
\usepackage{xcolor}
\usepackage{amsaddr}
\usepackage{amsmath,amsfonts,amssymb}
\usepackage{algorithm}
\usepackage{algpseudocode}
\usepackage{float}
\usepackage{upgreek}
\usepackage[version=4]{mhchem}
\usepackage{subcaption}
\usepackage{array}
\usepackage{bbm}

\usepackage{comment}

\DeclareMathOperator{\sign}{sign}
\newcommand{\s}{\mathbb{S}}

\newcommand{\R}{\mathbb{R}}
\newcommand{\N}{\mathbb{N}}

\newcommand{\Z}{\mathbb{Z}}

\newcommand{\rtwoD}{\widehat r_\mathrm{2D}}
\newcommand{\rTEM}{r_\mathrm{2D}}
\newcommand{\rthreeD}{r_\mathrm{3D}}

\newcommand{\softmax}{\sigma}
\newcommand{\randomField}{Z}
\newcommand{\roughness}{\gamma}

\newcommand{\minDiameter}{d_\mathrm{min}}
\newcommand{\maxDiameter}{d_\mathrm{max}}
\newcommand{\area}{A}
\newcommand{\elongation}{\eta}

\newcommand{\volume}{V}
\newcommand{\surfaceArea}{A_\mathrm{surf}}
\newcommand{\specificSurface}{S}
\newcommand{\sphericity}{\Phi}

\newcommand{\radiusSpherePacing}{r_\mathrm{sphere}}

\newcommand{\BN}{\mathrm{BN}}

\definecolor{ulmgruen}{rgb}{0.3372,0.6667,0.1098}
\definecolor{ulmorange}{rgb}{0.87,0.426,0.027}

\author{Lukas Fuchs$^1$, Kerstin Wein$^2$, Jens Friedland$^2$, Orkun Furat$^1$, Robert Güttel$^2$, Volker Schmidt$^1$}
\address{$^1$ Ulm University, Institute of Stochastics, Helmholtzstraße 18, 89069 Ulm, Germany}
\address{$^2$ Ulm University, Institute of Chemical Engineering, Albert-Einstein-Allee 11, 89069 Ulm, Germany}

\title[Stereological 3D modeling of catalyst particles]{Stereological  3D modeling of nano-scale catalyst particles using TEM projections}

\begin{document}

\begin{abstract}

Catalysis, particularly heterogeneous catalysis, is crucial in the chemical industry and energy storage. Approximately 80\% of all chemical products produced by heterogeneous catalysis are produced by solid catalysts, which are essential for the synthesizing of ammonia, methanol, and hydrocarbons. Despite extensive use, challenges in catalyst development remain, including enhancing selectivity, stability, and activity. These effective properties are influenced by the nanoscale morphology of the catalysts, whereby the size of the nanoparticles is only one key descriptor. To investigate the relationship between nanoparticle morphology and catalytic performance, a  comprehensive 3D analysis of  nano-scale catalyst particles is necessary. However, traditional imaging techniques for a representative recording of this size range, such as transmission electron microscopy (TEM), are mostly limited to 2D. Thus, in the present paper,  a stochastic 3D  model  is developed for a data-driven analysis of the  nanostructure  of catalyst particles. The calibration of this model is achieved using 2D TEM data from two different length scales, allowing for a statistically representative 3D modeling of catalyst particles. Furthermore, digital twins of catalyst particles can be drawn for the stochastic 3D model for  virtual materials testing, enhancing the understanding of the relationship between catalyst nanostructure and performance.

\smallskip
\noindent

\textbf{Keywords: Transmission electron microscopy, stereology, solid catalyst, simulated particle,  stochastic nanoscale modeling, random field, sphere packing,  generative adversarial network} 
\end{abstract}
\maketitle
\section{Introduction}
\noindent

Catalysis in general and heterogeneous catalysis in particular play important roles in the chemical industry as well as in chemical energy storage processes. In case of heterogeneous catalysis, approximately 80\% of all chemical products in the world are produced using solid catalysts~\cite{Hu2021}, where the large-scale synthesis of chemical energy storage molecules, such as ammonia, methanol or hydrocarbons,  relies completely on solid catalysts~\cite{refaat2011biodiesel,Rinaldi2009}. Even though heterogeneously catalyzed reactions are implemented on large industrial scales already for more than a century, major challenges remain for catalyst development, such as improved selectivity, stability, and activity. The workhorses for these catalytic parameters are supported solid catalysts, where catalytically active nanoparticles are supported on porous substrates in order to provide sufficient dispersion and stabilization of those nanoparticles. The catalytic performance -- expressed by descriptors such as activity and selectivity -- of such materials usually depends on their structural and geometrical properties, especially in the case of solid catalysts. The size of the active nanoparticles, for instance, affects significantly the activity and selectivity in various cases. In 1990 such a structure-property relationship was found in the \ce{Au}-based catalysts for \ce{CO} oxidation~\cite{HARUTA1993175} and is known as the particle size effect. More recently and also of interest for energy storage applications, a particle size effect has been reported in \cite{den2009origin} for the \ce{Co}-catalyzed Fischer-Tropsch reaction~\cite{den2009origin}, which was confirmed by other authors also for \ce{Fe}- and \ce{Ru}-based catalysts~\cite{PARK201084,CARBALLO2011102}. Interestingly, these results show a clear effect of the particle size on both the activity in \ce{CO} hydrogenation and selectivity in \ce{CH4} formation. In addition, the stability of the nanoparticles is dependent on their size, as was shown for thermal sintering~\cite{Cleays2014} and oxidation~\cite{WOLF20211014} under reaction conditions. Therefore, the design of sophisticated and well-defined solid materials are in the focus of research, with emphasis on controlling and stabilizing the nanoparticle structure. However, functional nanoparticle systems are characterized by a wide range of tuneable structural properties, which are defined by complex multidimensional probability distributions of descriptors, such as particle size, shape, and surface area~\cite{Frank2022}. Hence, tailoring of such nanoparticle systems for the use as solid catalyst requires the determination of the structural descriptors to establish the correlation to catalytic performance descriptors. The major challenges for respective material characterizations arise from the stochastic distribution and the experimental accessibility of the structural properties.

Transmission electron microscopy (TEM) is commonly used as a standard catalyst characterization technique. However, this technique often allows only for the acquisition of 2D (projection) images from the 3D samples, where the 2D projection data is typically simplified to an aggregated information, e.g., representing the mean particle size. Assuming that a statistically representative sample is measured, this value can be correlated to the catalytic performance. Therefore, the particle size is one key descriptor, as many properties of particulate products are influenced by their size distribution, and the analysis of particle systems was often reduced to univariate particle size distributions~\cite{leschonski1974a,leschonski1974b}. But, only for spherical particles, their size can be given by using a single structural descriptor, i.e., the diameter. 
We also remark that the derivation of univariate size distributions from multivariate distributions of particle descriptor vectors is easily possible while the reverse procedure, i.e., the derivation of multivariate  distributions from univariate ones, is an ill-posed problem due to lack of information regarding the correlation of components~\cite{Frank2022}. However, the nanoparticle size is not the only relevant descriptor determining the surface area of particles and their catalytic performance~\cite{Munirathinam2018}. Instead, it has been shown that the structure of the surface, e.g., the presence of kink and edge sites~\cite{Illner2023}, governs the mechanism and kinetics of the reactions taking place~\cite{Kläger2023}. Non-spherical nanoparticles have various surface descriptors, where not only the particle diameter, but additional descriptors such as sphericity, are needed to accurately characterize the particle surface. Thus, 3D information on nanoparticle morphology is highly desired to understand the relationship between catalyst nanostructure and catalytic performance. For this purpose, 3D TEM is a valuable option~\cite{strass2021bifunctional,ERSEN20071088} involving the collection of several 2D TEM images at different angles, followed by the reconstruction of a 3D representation of the object under consideration. However, this approach is often too costly in terms of time and resources to be used even for a single particle, and much less for a representative number of particles. 

The present paper aims to overcome this issue by stereologically fitting a stochastic 3D particle model using 2D image data to generate statistically similar digital twins of real  nanoparticles, see~\cite{2D-3D_approach,2D-3D_approach_2,SliceGan,fuchs2024generating} for related stereological modeling approaches. The realization of such a 3D model can then be used for 3D nanostructure analysis as well as for the investigation of effective properties  through numerical simulations~\cite{hofmann2020electro,weber2017simulation}.
The proposed 3D model consists of two parts. First, rough 3D shapes of particles are modeled. Then, a second modeling component adds finer surface features, generating particles with nanometer-scale details. This two-stage modeling approach leverages measurement techniques at different length scales: a coarser resolution captures the size and overall shape  of particles for high representativity, while a finer resolution captures detailed surface features for high accuracy. Specifically, a TEM image with a pixel size of 0.678 nm is used to acquire a representative quantity of rough particle shapes, and a TEM image with a pixel size of 0.102 nm is used to capture the fine surface features of a limited number of particles.
The model used for the description of rough particle hulls consists of a random field on the unit sphere in the three-dimensional Euclidean space $\R^3$, 
based on a spherical harmonics  representation. While there are several attempts to stereologically fit spherical harmonics~\cite{Stereological_SH_1,Stereological_SH_2}, these attempts focus on planar sections through particles rather than projections. In contrast, the present paper proposes a novel approach to fit a random field model on the unit sphere in  $\R^3$, defined through random linear combinations of spherical harmonics, solely using 2D projection data.
To achieve this, two neural networks are introduced: one for generating random linear combination coefficients (the so-called generator) and one for model fitting (the so-called discriminator). This neural network-based approach allows for efficient model fitting within a generative adversarial network (GAN) framework~\cite{GAN,GAN_overview}.
The subsequent modeling of the fine surface features is achieved through an overlapping sphere packing algorithm~\cite{FB_original,FB_overview} and a morphological closing operation~\cite{MorphologicalClosing}. If correctly calibrated, by means of this method it is possible to generate shape outlines such as those observed in high resolution TEM image data of the catalyst particles.

The rest of the present paper is organized as follows. In Section~\ref{sec.tio.two}, the materials and methods used in this paper are explained.
First,
in Section~\ref{sec.two.one}, the considered particles, their synthesis, and imaging are described in detail. Then, in Section~\ref{sec.two.sta}, the coarser particle hull model and 
the fine surface feature model are introduced. Afterward, in Sections~\ref{sec:stereological_hull_fitting} and~\ref{sec:sphere_packing_fitting},
methods for the stereological calibration of both 3D models by means of 2D TEM data are stated.   In Section~\ref{sec:results}, the whole modeling procedure is evaluated and predictions of 3D particle morphologies are presented, utilizing realizations of the stochastic multi-scale model. Finally, Section~\ref{sec.tio.fou} concludes and provides an outlook to possible future research.

\section{Materials and methods}\label{sec.tio.two}
\subsection{Synthesis, TEM  imaging and image processing}\label{sec.two.one}

Colloidal cobalt oxide (\ce{Co3O4}) nanoparticles were prepared via solvothermal treatment. Therefore, polyvinylpyrrolidone (PVP, MW 1,300,000, high purity grade, AMRESCO) is dissolved in ethanol (EtOH, 96vol\%, VWR CHEMICALS) in a first step. As soon as the PVP is molecularly dissolved, cobalt nitrate (Cobalt(II) nitrate hexahydrate, 97.7\%, ALFA AESAR) is added to the solution and stirred until a homogeneous solution is obtained. The solution is then transferred into a Teflon liner and placed in an autoclave. The solvothermal treatment is carried out at 180 °C for 3 h leading to a suspension containing \ce{Co3O4} nanoparticles. From this base-suspension (stabilized as-made \ce{Co3O4}/PVP/EtOH solution) the samples were prepared. Therefore, one droplet of the base-suspension was mixed with EtOH and homogenized via ultrasonication, resulting in the TEM-suspension.

Low- and high-resolution TEM image data were collected. The low-resolution image captures a statistically representative sample of the particle  sizes and shapes, while the high-resolution image is focusing on detailed surface features. This two-stage approach ensures that the dataset encompasses both the overall morphology as well as finer surface details of the particles, providing a comprehensive basis for  model calibration considered in Sections~\ref{sec:stereological_hull_fitting} and~\ref{sec:sphere_packing_fitting}. 

\begin{figure}[ht!]
    \begin{minipage}{.22\textwidth}
    \centering
    \includegraphics[width=\textwidth]{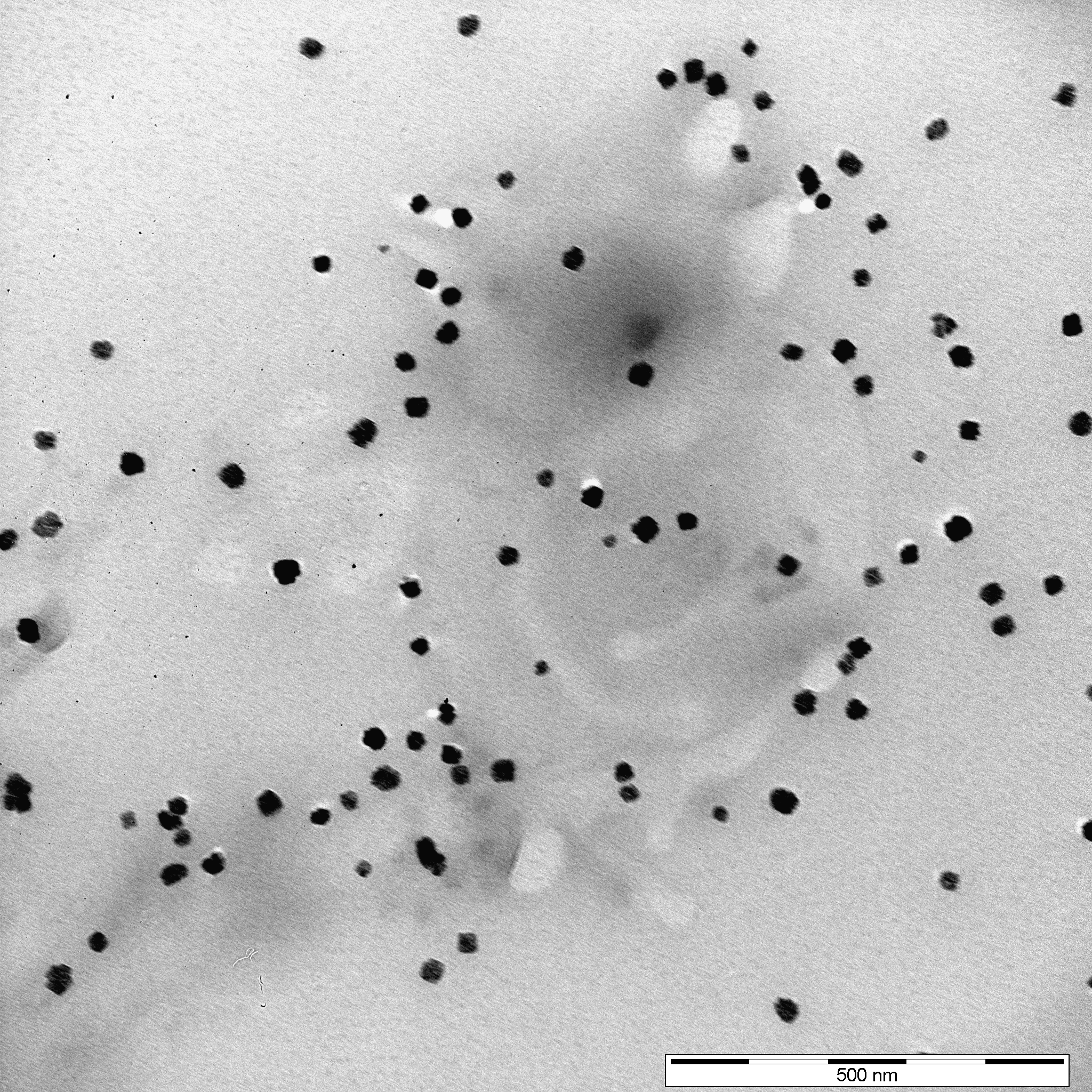}\\(a)
    \end{minipage}
    \begin{minipage}{.22\textwidth}
    \centering
    \includegraphics[width=\textwidth]{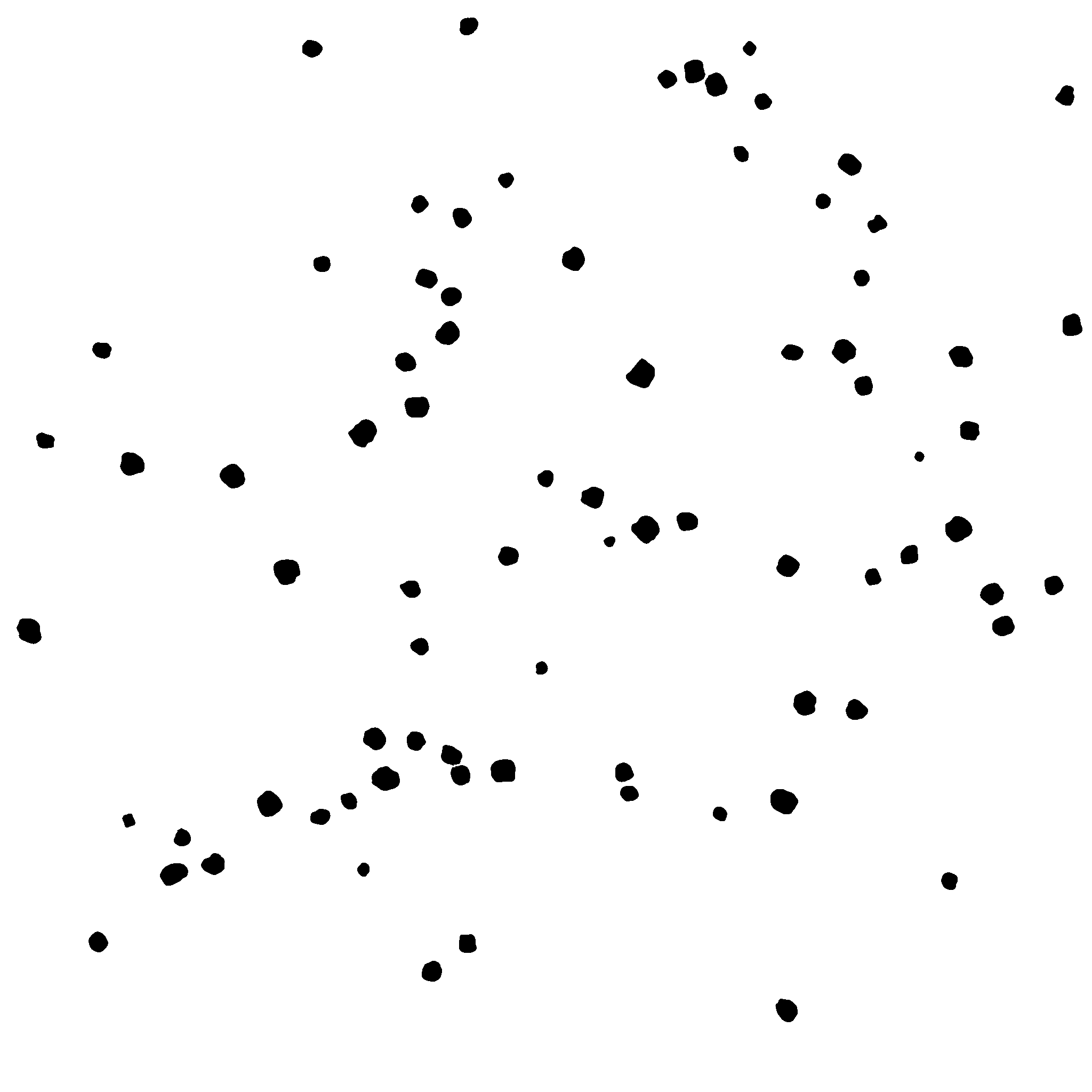}\\(b)
    \end{minipage}
    \begin{minipage}{.22\textwidth}
    \centering
    \includegraphics[width=\textwidth]{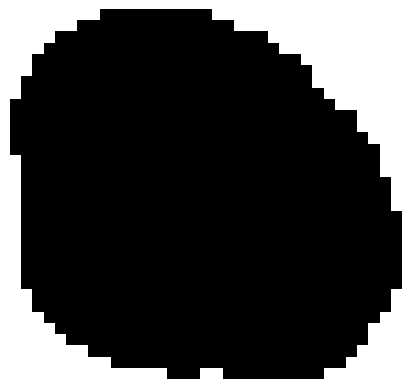}\\(c)
    \end{minipage}
    \begin{minipage}{.22\textwidth}
    \centering
    \includegraphics[width=\textwidth]{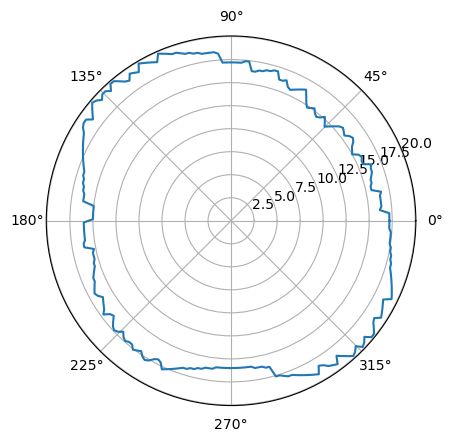}\\(d)
    \end{minipage}
    \caption{Data preprocessing steps: (a) Low-resolution TEM image (projection) of particles. (b) Segmentation of image data  into individual non-overlapping particles. (c) Magnification of a single segmented particle projection. (d) Radial representation of the particle outline. }
    \label{fig:preprocessing}
\end{figure}

For the low resolution TEM measurements, one droplet of the TEM suspension was applied on a graphitized copper grid (300 mesh) and subsequently dried in the desiccator. The TEM images have been obtained by means of a JOEL 1400 microscope operating at 120 kV, see  Figure~\ref{fig:preprocessing}a.
Image processing, i.e., the extraction of non-overlapping particles is done in two steps. First,   the particle phase is segmented from the background using thresholding~\cite{Otsu}. Subsequently, individual particles are identified by manual removal of overlapping particles. The outcome of these procedures is depicted in Figure~\ref{fig:preprocessing}b. 
The magnification of a single segmented particle projection in shown in Figure~\ref{fig:preprocessing}c, together with the radial representation of the particle outline in Figure~\ref{fig:preprocessing}d, see 
Sections~\ref{sec:outer_hull_model} and~\ref{sec:stereological_hull_fitting}    for  details.

\begin{figure}[ht]
\centering
     \begin{subfigure}{0.22\textwidth}
     \centering
         \includegraphics[height = \textwidth]{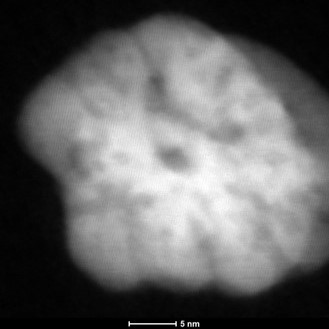}
         \caption{}
     \end{subfigure}
     \hspace{1cm}
     \begin{subfigure}{0.22\textwidth}
     \centering
         \includegraphics[height = \textwidth]{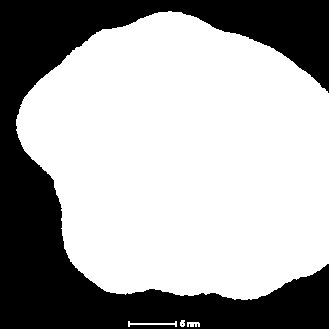}
     \caption{}
     \end{subfigure}
     \hspace{1cm}
     \begin{subfigure}{0.22\textwidth}
     \centering
         \includegraphics[height = \textwidth]{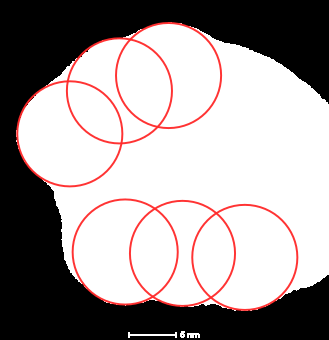}
     \caption{}
     \end{subfigure}
     
    \caption{(a) High-resolution TEM image of a single particle. (b) Segmentation of the (projected) particle outline. (c) Sphere-based representation  of surface roughness.}
    \label{fig:high_resolved}
\end{figure}

Furthermore, high-resolution TEM images have been acquired at the Institute of Material Science Electron Microscopy of Ulm University using a Thermofisher 200X (scanning) TEM operated at 200 kV, see  Figure~\ref{fig:high_resolved}a. For these high-resolution TEM measurements, the suspension was drop-casted on holey carbon TEM support grids. Various imaging modes were applied, such as conventional bright-field HRTEM imaging and acquisitions of selected area electron diffraction patterns. The segmentation of the (projected) particle outline is shown  in Figure~\ref{fig:high_resolved}b, whereas  Figure~\ref{fig:high_resolved}c  illustrates the sphere-based representation  of surface roughness, see   Section~\ref{sec:sphere_packing_model} for details.

\subsection{Stochastic multi-scale  modeling approach}
\label{sec.two.sta}

In this section, we introduce the  stochastic multi-scale  model which will be exploited to simulate the 3D morphology of nanoparticles, where this model generates 3D digital twins of real nanoparticles based on 2D image data. The model consists of  two parts. First,  a  corse particle model is stated, capturing the sizes and  overall shapes of particles. It will be called the particle hull model in the following. Then,   a second stochastic model is presented, which adds detailed surface features to simulated particle hulls. This is done by representing the surface by means of overlapping spheres, which mimic the surface details seen in the high-resolution image data.

\subsubsection{Particle hull model}\label{sec:outer_hull_model}

We assume that the 3D particles considered in this paper are star-shaped sets  $P\subset\R^3$. This means that there is a point $x\in P$, the so-called star-point, such that for each $y\in P$ the entire segment $[x,y]$ of points between $x$ and $y$ is contained in the set $P$, i.e., $x+t(y-x)\in P$ for each $t\in[0,1]$~\cite{preparata85}. Moreover, without loss of generality, we  assume that the star-point is located at the origin $o\in\R^3$.

The particle hull is then modeled by a random field $Z=\{Z(x), x\in\s\}$ on the unit sphere $\s=\{x\in\R^3: |x|=1\}$. By considering a parametrization of $\s$  by means of pairs of azimuthal and polar angles $(\phi, \theta)\in[0, 2 \pi) \times [0, \pi)$, the particle hull model is given as a family $Z=\{Z(\phi,\theta)\}$ of random variables
$Z(\phi,\theta)$, where $\phi\in[0, 2 \pi), \theta\in[0, \pi)$.   Realizations of this model are 3D radius functions $\rthreeD\colon [0, 2 \pi) \times [0, \pi) \to \R$ representing digital twins of particle hulls, see Figure~\ref{fig:preprocessing}d. More precisely, the proposed random field model $\randomField$ is defined as a random linear combination of real-valued basis functions,  so-called  spherical harmonics, $Y_{\ell m}\colon [0, 2\pi) \times [0, \pi) \to \R$ with $m\in \{-\ell,-\ell+1,\ldots,\ell\},~\ell \in \{0,1,\ldots\}$, which are given by
\begin{align}\label{def.sph.har}
    Y_{\ell m}(\phi,\theta)=\begin{cases}
 \displaystyle \sqrt{2}\sqrt{\frac{2\ell +1}{4 \pi}\frac{(\ell -|m|)!}{(\ell+|m|)!}} P^{|m|}_\ell(\cos\theta)\sin(|m|\phi), & \text{if } m <  0, \\
 \displaystyle
  \sqrt{\frac{2\ell +1}{4 \pi}} P^{m}_\ell (\cos\theta),  & \text{if } m = 0,\\
  \displaystyle\sqrt{2}\sqrt{\frac{2\ell +1}{4 \pi}\frac{(\ell-m)!}{(\ell+m)!}} P^{m}_\ell (\cos\theta)\cos(m\phi), & \text{if } m > 0,
\end{cases}
\end{align}
where $P^{m}_\ell \colon [0,1] \to [0,1]$ are Legendre polynomials~\cite{SH}. Note that the functions $Y_{\ell m}$ given in Eq.~\eqref{def.sph.har} are widely used in particle modeling~\cite{furat2021artificial,feinauer2015structural} and computer graphics~\cite{ComputerGraphics} due to their ability of representing a large range of radius functions on the unit sphere. In fact, for unbounded order $\ell$, linear combinations of them can represent all reasonable particle surfaces\footnote{The family of spherical harmonics forms a basis of the space of square-integrable functions on the unit sphere.}. However, spherical harmonics of increasing order mainly contribute to capture more detailed and rougher surface features, which are often not observable in data. Therefore, in applications, spherical harmonics of too high orders are neglected due to computational efficiency and the discrete nature of measured image data. In this paper, spherical harmonics up to an order of $\ell= 6$ are considered. A realization $\rthreeD\colon [0, 2 \pi) \times [0, \pi) \to \R$ of the random field model $Z$  is thus given by
\begin{align}\label{rea.ran.fie}
    \rthreeD(\phi,\theta) = \sum_{\ell=0}^6\sum_{m=-\ell}^l a_{\ell m}Y_{\ell m}(\phi, \theta),
\end{align}
for each $(\phi,\theta)\in[0, 2 \pi)\times [0, \pi)$, where
the prefactors $a_{\ell m}$ are referred to as spherical harmonics coefficients. The stochastic modeling of these coefficients results in the particle hull model $\randomField=\{Z(\phi,\theta)\}$ with
\begin{align}\label{zet.lin.com}
    \randomField(\phi,\theta) = \sum_{\ell=0}^6\sum_{m=-\ell}^l f_{\ell m}(X)Y_{\ell m}(\phi, \theta)
\end{align}
for each $(\phi,\theta)\in[0, 2 \pi)\times [0, \pi)$, where $X=(X_1,\ldots,X_{49})$ is a 49-dimensional random vector of  independent standard normal distributed random variables, and  $f_{\ell m} \colon \R^{49} \to \R$ is some spherical harmonics coefficient function,  for any $\ell \in \{0,1,\ldots,6\}, m\in \{-\ell,-\ell+1,\ldots,\ell\}$. Note that the dimension $49$ corresponds to the number of spherical harmonics up to order 6, i.e., $49 = \sum_{\ell=0}^6(2l+1)=  \sum_{\ell=0}^6\sum_{m=-\ell}^l 1$.
The coefficient functions $f_{\ell m}$ considered in Eq.~\eqref{zet.lin.com} will be represented by a high-parametric fully connected neural network, see Section~\ref{sec:stereological_hull_fitting}. On the one hand, this enables the modeling of random  coefficients $f_{\ell m}(X)$  with complex correlation structure and, on the other hand, allows for an efficient fitting of the particle hull model $\randomField$ to data,  through gradient descent-based neural network training, see Section~\ref{sec:stereological_hull_fitting}.

\subsubsection{Sphere packing model}\label{sec:sphere_packing_model}  
In the previous section, a stochastic model for the sizes and overall shapes of particles was introduced. Later on, in Section~\ref{sec:stereological_hull_fitting}, we show how this model can be
 calibrated using a large dataset of low-resolution particle projections, making it suitable for representatively describing rough particle morphologies. However, the particle hull model introduced in Section~\ref{sec:outer_hull_model}  
 is not suitable for depicting fine structural features of particle surfaces, as this would require a significant increase in evaluation points and measurement resolution, which is costly in both time and resources. To address this issue, a second model is employed, to add detailed surface features to rough particle hulls. The calibration of this second model will be explained in Section~\ref{sec:sphere_packing_fitting} below, where the resulting fine features match those of particles measured by high-resolution imaging, see Figure~\ref{fig:high_resolved}a. Remarkably, this approach drastically reduces the need for numerous cost-intensive high-resolution measurements.

\begin{figure}[ht]
\centering
     \begin{subfigure}{0.22\textwidth}
         \centering
         \includegraphics[height =\textwidth]{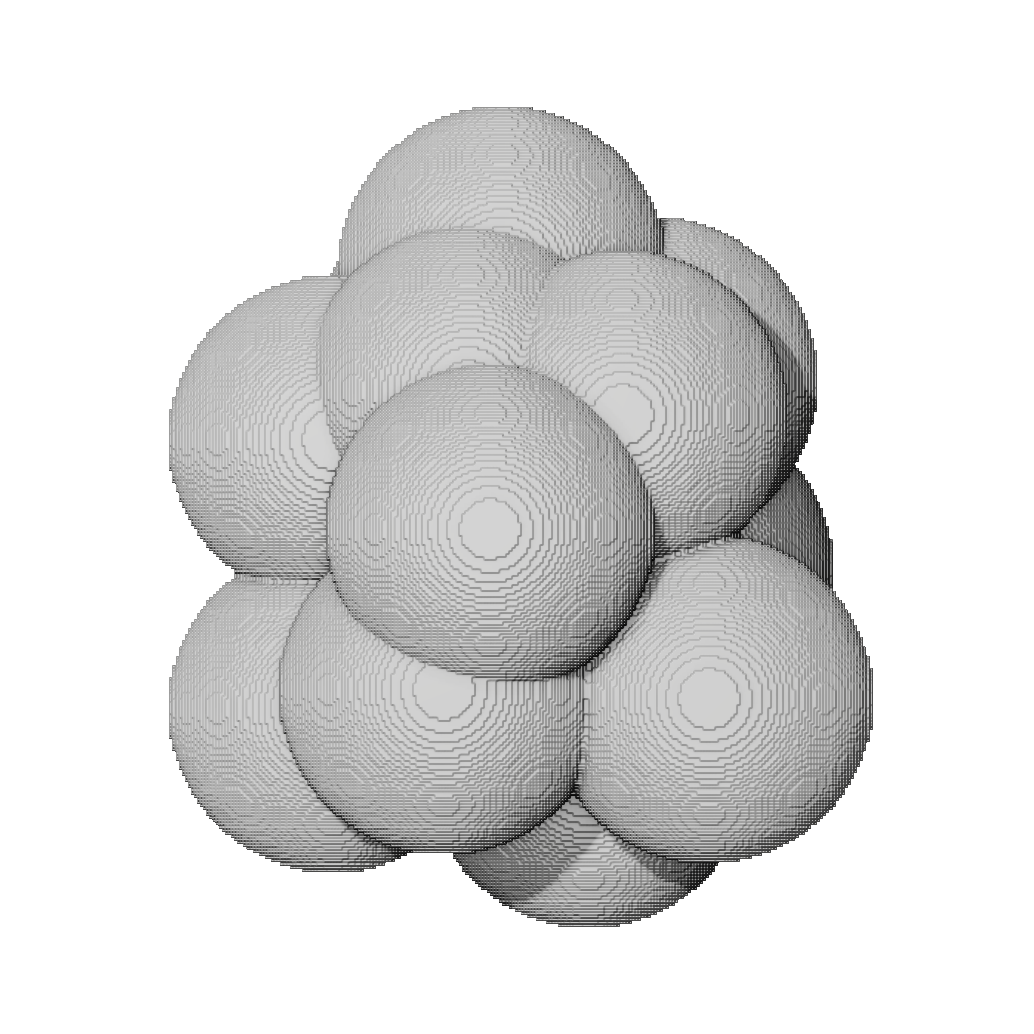}
     \caption{}
     \end{subfigure}
     \hspace{0.1cm}
     \begin{subfigure}{0.22\textwidth}
         \centering
         \includegraphics[height =\textwidth]{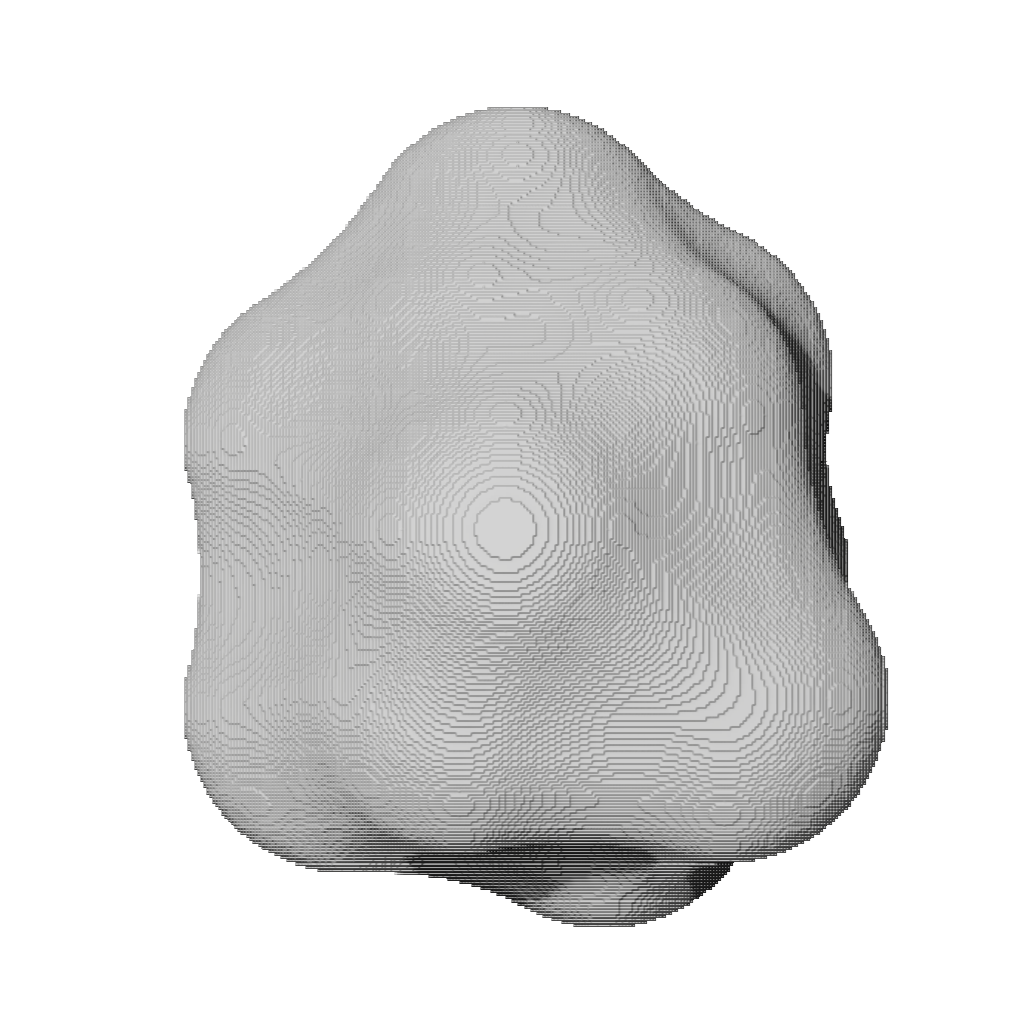}
     \caption{}
     \end{subfigure}
     \hspace{0.1cm}
     \begin{subfigure}{0.22\textwidth}
     \centering
         \includegraphics[height = \textwidth]{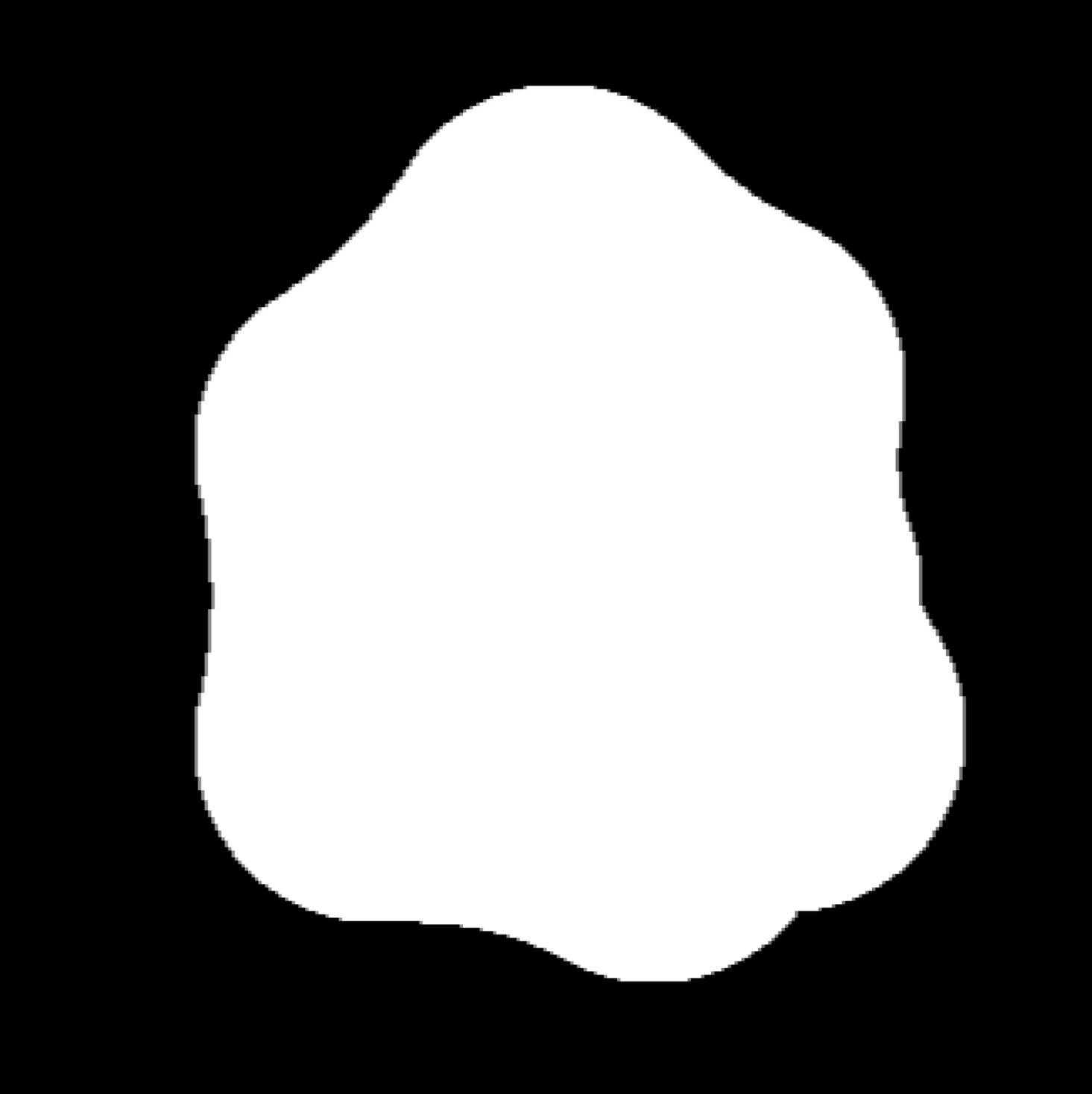}
     \caption{}
     \end{subfigure}
     \begin{subfigure}{0.22\textwidth}
         \includegraphics[height =\textwidth]{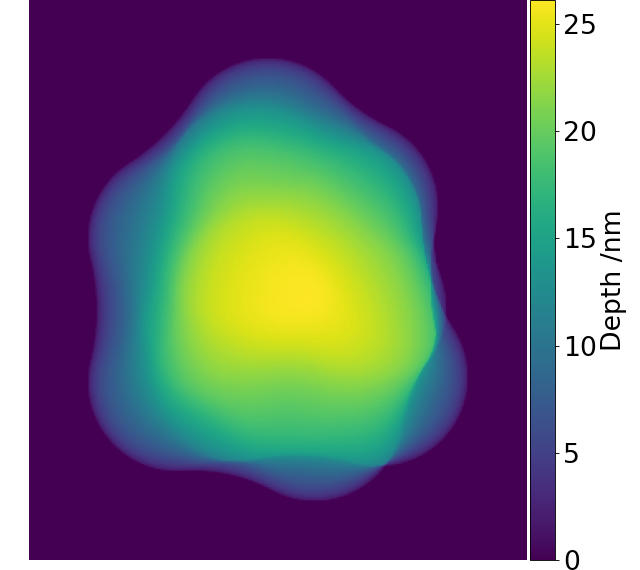}
     \caption{}
     \end{subfigure}
    \caption{ (a)  3D overlapping sphere packing. (b) Simulated 3D particle after applying morphological closing. (c) Outline of  simulated particle. (d) Thickness profile of simulated 3D particle. 
    }
    \label{fig:high_resolved_2}
\end{figure}

The basic idea of this second modeling approach is the fact that the fine 2D surface features observable in Figure~\ref{fig:high_resolved}a  can be accurately represented by overlapping circles of equal size, i.e., projections of mono-dispersed spheres, see Figure~\ref{fig:high_resolved}c. This motivates the use of an overlapping sphere packing  to model fine structural features of  3D particle surfaces adequately.
More precisely,
 fine surface feature modeling is achieved using an overlapping sphere packing algorithm~\cite{FB_original,FB_overview}  followed by a morphological closing operation~\cite{MorphologicalClosing}. Specifically, for a given integer $n>1$ and some radius $\radiusSpherePacing>0$, a simplified forced bias algorithm packs $n$ overlapping spheres of radius $\radiusSpherePacing$ within a particle hull $P\subset\R^3$, which was  previously  generated by the random field model described  in Section~\ref{sec:outer_hull_model},   see Figure~\ref{fig:high_resolved_2}a. 
Then, in a second step, a morphological closing operation closes gaps between the spheres in order to yield a projected particle surface as the one observed in high-resolution image data, compare Figures~\ref{fig:high_resolved}b and~\ref{fig:high_resolved_2}c.

Sphere packing within a 3D particle hull $P\subset\R^3$ is done in the following way. First,  by drawing $n$ points (independently of each other) from the uniform distribution on the set $P$,  a point pattern in $P$ is generated which 
corresponds  to the initial configuration of sphere centers.   These points are then greedily moved within the set $P$ to minimize the  overlap of their associated spheres. More precisely, for each $p \in P$, the closest point $p^\prime \in P$  such that $p^\prime \neq p$, and the closest point $x \in P^\mathrm{c}=\R^3\setminus P$  outside the set $P$   are computed. Finally, the point $p\in P$ is  moved as follows:
\begin{align}
p \mapsto \begin{cases}
p + \frac12(p-x) ,& \text{if } |p-x| < \radiusSpherePacing,\\
p + \frac14(p-p') ,& \text{if } |p-x| \geq \radiusSpherePacing \text{ and } |p-p'|<2\radiusSpherePacing,\\
p ,& \text{otherwise,}
\end{cases}
\end{align}
where $|x|=\sqrt{x_1^2+x_2^2+x_3^2}$ denotes the Euclidean norm of $x=(x_1,x_2,x_3)\in\R^3$, i.e.,  the length of the vector $x$. 
The procedure stops after 10,000 iterations or when a packing of non-overlapping spheres inside $P$ is achieved earlier, meaning that no points are moved anymore.
The resulting (final) configuration of sphere centers will be denoted by $p_1,\ldots, p_n\in P$, i.e., $B=\{x\in\R^3: |x-p_i|\le \radiusSpherePacing \text{ for some } i\in\{1,\ldots,n\}\}$ is the union set of the $n$ packed spheres.

In a second step, a morphological closing~\cite{MorphologicalClosing} is applied to the  sphere packing $B\subset\R^3$  described above, in order to fill the gaps between spheres and to yield a smoother surface, see  Figures~\ref{fig:high_resolved_2}a
 and~\ref{fig:high_resolved_2}b.  In this context, for any fixed $\delta\ge 0$, the morphological closing $c_\delta \colon \mathcal{P}(\R^3) \to \mathcal{P}( \R^3)$  is defined as superposition of the erosion $e_\delta \colon \mathcal{P}(\R^3) \to \mathcal{P}( \R^3)$ and the dilation $d_\delta \colon \mathcal{P}(\R^3) \to \mathcal{P}( \R^3)$, where $\mathcal{P}(\R^3)$ denotes the family of all subsets of $\R^3$,  and, for each set $B\in\mathcal{P}(\R^3)$, it holds that
\begin{align}
d_\delta(B)=& \{x \in \R^3 \colon \text{ there is a } y \in B \text{ such that } |x-y|\leq \delta\}\\
e_\delta(B)=&\{x \in B \colon \text{ there is no } y \in \R^3 \setminus B \text{ such that } |x-y|<\delta\}.
\end{align}
Thus, $
c_\delta(B)=e_\delta(d_\delta(B))$ for each $B\in\mathcal{P}(\R^3)$, 
where $\delta \in [0, \infty)$ is a parameter that defines the magnitude of the morphological closing. In particular, a value of $\delta = 0$ results in the original sphere packing, i.e., $c_0(B)=B$, while $\displaystyle\lim_{\delta \to \infty} c_\delta(B)$ is the convex hull of $B$, where the volume of the set  $c_\delta(B)$ increases monotonously with $\delta\ge 0$.

In total, the sphere packing model described above  has three tunable parameters, namely, the number $n$ of placed spheres, their radius $\radiusSpherePacing$,  and the magnitude $\delta$ of the applied morphological closing.  An optimization method to best select the values of these parameters will be discussed in Section~\ref{sec:sphere_packing_fitting} below.

\subsection{Sterological fitting of the particle hull model}\label{sec:stereological_hull_fitting}

The electron microscopy image data described in Section~\ref{sec.two.one}  is used to stereologically calibrate the  stochastic multi-scale  model stated in Section~\ref{sec.two.sta}. 
This is done in two steps. First, in this section, the low-resolution TEM image is exploited to calibrate a 3D model of corse particles, capturing their sizes and  overall shapes.  Then, in Section~\ref{sec:sphere_packing_fitting}, the high-resolution TEM image is utilized to calibrate the sphere model, in order to add detailed surface features to particle hulls.

\subsubsection{Radial representations of particle projections.}\label{sec:radial_rep}

Recall that in Section~\ref{sec:outer_hull_model} we assumed that the 3D particles considered in this paper are star-shaped sets in  $\R^3$.
From this it follows that their projections are star-shaped sets as well, but now in the two-dimensional Euclidean plane $\R^2$, i.e., the 2D projections of the 3D particles have radial representations in $\R^2$.

To  calibrate the particle hull model introduced in  Section~\ref{sec:outer_hull_model} by means of 
 the  projections of 3D particles extracted from low-resolution TEM image data, see Figure~\ref{fig:preprocessing}c,  the following radial representation of particle projections is chosen to compare the  projections computed from TEM data with those of simulated particles.  To simplify the notation we assume without loss of generality that the pixel size of TEM data is equal to 1. Furthermore,  for the  projection of a particle extracted from  TEM data, the function $\zeta \colon \Z^2 \to \{0,1\}$ is considered, where 
\begin{align}\label{def.fun.zet}
    \zeta(x)= \begin{cases}
        1,~\text{pixel $x$ belongs to the particle projection,}\\
        0,~\text{else,}
    \end{cases}
\end{align}
for each $x\in\Z^2$, and  $\Z=\{\ldots,-1,0,1,\ldots\}$ denotes the set of all integers. Next,
the function $\zeta$ defined in Eq.\eqref{def.fun.zet} is extended to the (continuous) Euclidean plane $\R^2$ by setting the values of $\zeta:\R^2\to\{0,1\}$ for each  $y\in \R^2$ to $\zeta(y)=\zeta(x)$, where $x\in\Z^2$ arises from $y$ by componentwise rounding to the nearest integer. The radial representation $\rTEM\colon [0,2\pi) \to [0,\infty)$ of $\zeta$ is then given by
\begin{align}
    \rTEM (\phi) = \sup \left \{r \ge 0 \colon \zeta \Bigl(r \begin{pmatrix}\cos(\phi)\\\sin(\phi) \end{pmatrix}+c \Bigr)=1\right\},  \label{eq:rtwod_measured}
\end{align}
for each $\phi \in [0,2\pi)$, where $c\in \R^2$ is the center of mass of the particle projection\footnote{Note that in implementations the supremum can be treated as maximum.}. Note that this procedure mathematically defines the  projected outline   of star-shaped 3D  particles (with projected star-point $c$), see also Figures~\ref{fig:preprocessing}c and~\ref{fig:preprocessing}d.

In order to be comparable with the radius function $\rTEM$ given in Eq.~\eqref{eq:rtwod_measured} for measured TEM  data, the radius function  $\rthreeD$ of simulated outer hulls given in Eq.~\eqref{rea.ran.fie}, i.e., realizations of the particle hull model $Z$ introduced in Section~\ref{sec:outer_hull_model}, has  to be virtually projected onto a 2D plane. In particular, such a projection $\widetilde{r}_{\mathrm{2D}}:[0,2\pi) \to [0,\infty)$ of  $\rthreeD$  in $(0,0,1)$-direction is given by
\begin{align}\label{pro.rad.fun}
    \widetilde{r}_{\mathrm{2D}}(\phi) = \max(\{ \rthreeD(\phi, \theta) \sin(\theta) \colon \theta \in [0, \pi)\}),
\end{align}
for each $\phi \in [0,2\pi)$.
Here,  $\sin(\theta)$ accounts for the fact that values of $\rthreeD(\phi, \theta)$ for angles $\theta\in[0,\pi)$ close to 0 or $\pi$, which represent directions nearly parallel to the projection direction, contribute less to the size and shape of the projection.
However, due to computational efficiency, the maximum in Eq.~\eqref{pro.rad.fun} will be determined with respect to some discretized version of the interval $[0,\pi)$, i.e., only angles $\theta\in \Theta=\{0, \frac{\pi}{8},\ldots, \frac{7\pi}{8}\}$ are considered. Furthermore, to ensure differentiability of the projection procedure, and thus enabling efficient model fitting by means of gradient descent algorithms,  a modified (differentiable) approximation $\softmax \colon \mathcal{P}_0(\R) \to \R $ of the  maximum function in Eq.~\eqref{pro.rad.fun}  is considered, which is given by
\begin{align}
    \softmax(M) = \sum_{m \in M} \frac{m e^{m}}{\sum_{m' \in M} e^{m'}} 
    = \left\langle \begin{pmatrix} m_1  \\ \vdots\\ m_n\end{pmatrix}, 
    \begin{pmatrix} \frac{e^{m_1}}{\sum_{m' \in M} e^{m'}} \\  \vdots \\ \frac{e^{m_n}}{\sum_{m' \in M} e^{m'}}  \end{pmatrix} \right\rangle 
    \approx \left\langle \begin{pmatrix}  m_1  \\ \vdots \\ m_n\end{pmatrix}, 
    \begin{pmatrix}   \mathbbm{1}_{m_1 = \max M}  \\ \vdots \\ \mathbbm{1}_{m_n = \max M} \end{pmatrix} \right\rangle 
    = \max M,\label{def.sof.max}
\end{align}
for each $M\in \mathcal{P}_0(\R)$, 
where $\mathcal{P}_0(\R)$ is the family of all finite (non-empty)  subsets of $\R$,   $(m_1, \ldots, m_n)$ is a fixed vector representation of the elements of $M$, $\mathbbm{1}$ is the indicator function, i.e., $\mathbbm{1}_{m_i = \max M}=1$ if ${m_i = \max M}$ is true and   $\mathbbm{1}_{m_i = \max M}=0$ if ${m_i \not= \max M}$,  and $\langle \cdot,\cdot\rangle$ denotes the inner product. Note that the differentiable approximation $\sigma(M)$ of $\max M$ given in Eq.~\eqref{def.sof.max} relies on the so-called softmax function as an approximation of the indicator function~\cite{pmlr-v70-asadi17a}. Thus, in the following, 
the approximate projected radius function $\rtwoD:[0,2\pi)\to[0,\infty)$ of $\rthreeD$ will be considered, which is given by
\begin{align}
    \rtwoD(\phi) = \softmax(\{\rthreeD(\phi,\theta) \sin(\theta)  \colon \theta \in \Theta \}),\label{eq:projection}
\end{align}
for each $\phi \in [0,2\pi)$.

To achieve computational efficiency in the subsequent fitting procedure of the particle hull model and to mitigate overfitting due to the discrete nature of measured particles,  a discretized version of the radius function $\rtwoD$ given in 
Eq~\eqref{eq:projection} 
is considered. Specifically, the radius functions $\rTEM$ and  $\rtwoD$ (either from the radial representation for a segmented particle in Eq.~\eqref{eq:rtwod_measured} or from projected realizations of the particle hull model $\randomField$ as defined in Eq~\eqref{eq:projection}) are associated with a sequence of its values evaluated on an equidistant grid of $\kappa =16$ angles: $\rTEM(\phi_0), \ldots, \rTEM(\phi_{\kappa-1})$ and $\rtwoD(\phi_0), \ldots, \rtwoD(\phi_{\kappa-1})$, where $\phi_i = \frac{2 \pi i}{\kappa}$ for $i\in \{0, \ldots, \kappa-1\}$. This approach reduces the resolution of the radius function to eliminate rough surface artifacts caused by the pixel-based representation of TEM images, like those illustrated in Figure~\ref{fig:preprocessing}d.

With a representation established that allows both  simulated particle hulls and measured particles to be expressed in the same way, i.e., through their radius functions $r_{\mathrm{2D}}$ and $\rtwoD$, enabling their comparison, the next section introduces the first part of a neural network approach,  the so-called generator, for functions $f\colon \mathbb{R}^{49} \to \mathbb{R}^{49}$ used to compute spherical harmonic coefficients within the particle hull model, as shown in Eq.~\eqref{zet.lin.com}, where 
\begin{equation}\label{def.vec.eff}
f=(f_{\ell m}, \ell \in \{0,1,\ldots,6\}, m\in \{-\ell,-\ell+1,\ldots,\ell\}).
\end{equation}

\subsubsection{Architecture of the generator network}\label{sec:generator_architecture}
Recall that the vector-valued function $f$ in Eq.~\eqref{def.vec.eff} defines the particle hull model by mapping the random vector $X=(X_1,\ldots,X_{49})$  of  independent standard normal distributed random variables to the random vector $f(X)$ of spherical harmonic coefficients, where the function $f$ will be implemented as a neural network~\cite{aggarwal2018neural}. Thus, $f$ can be expressed as a superposition of simple parametric functions, commonly referred to as layers. By employing layers that are almost everywhere differentiable, the parameters of $f$ can be efficiently calibrated through gradient descent, leading to an efficient calibration of the particle hull model.

\begin{figure}[H]
    \centering
    \includegraphics[width=0.5\linewidth]{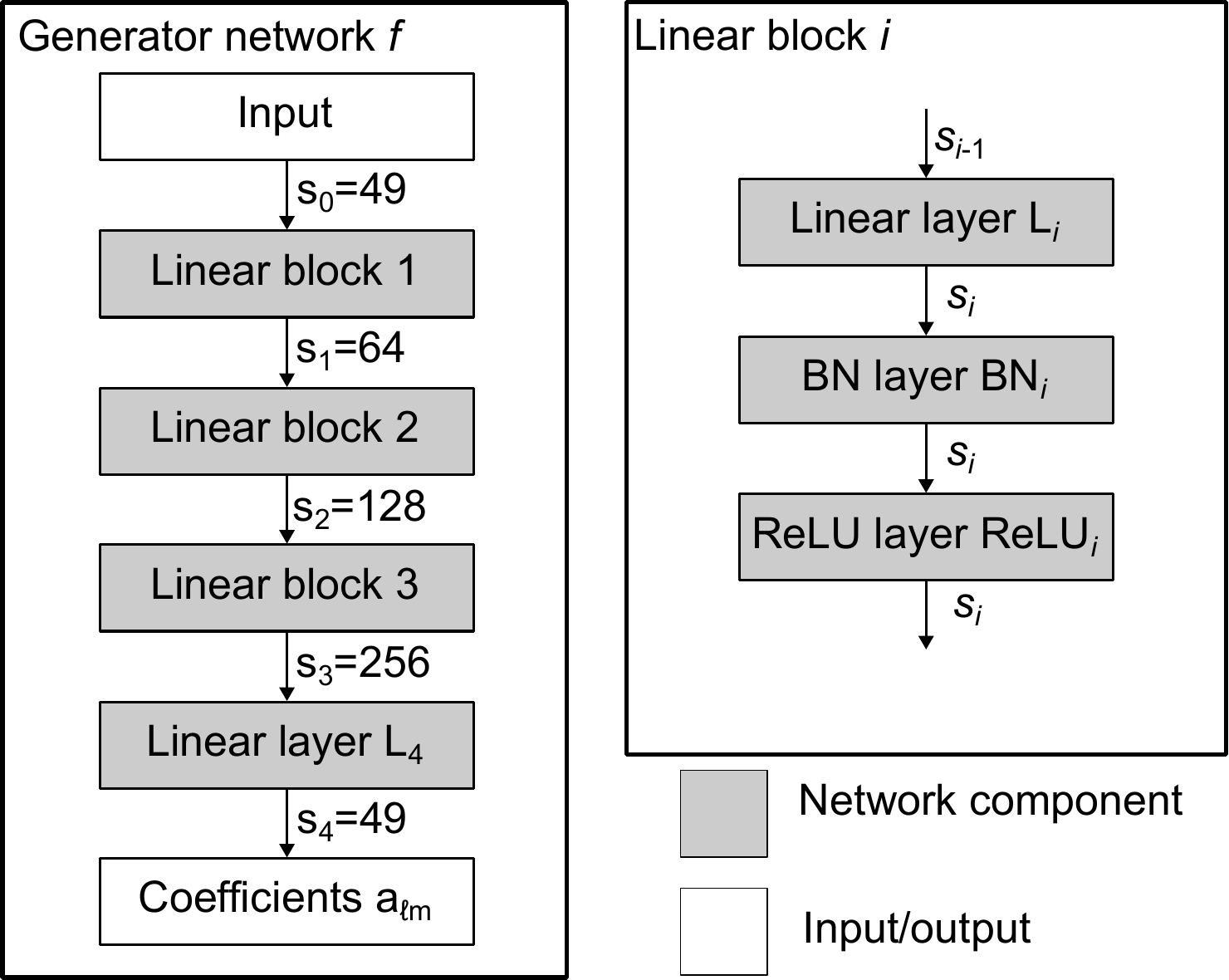}
    \caption{Architecture of the generator  network  for computing realizations $a_{\ell m}$ of the random spherical harmonics coefficients $f_{\ell m}(X)$. Left: General network architecture with output sizes $s_0,\ldots,s_4$ assigned to the down arrows. Right: Inner structure of linear blocks.}
    \label{fig:generator_architecture}
\end{figure}

More precisely, 
the parametric function $f\colon \R^{49} \to \R^{49}$ has the form
\begin{align}\label{def.neu.net}
    f = L_4 \circ \text{ReLU}_3 \circ \BN_{3} \circ L_{3} \circ \cdots \circ \text{ReLU}_1 \circ \BN_1 \circ L_1 ,
\end{align}
where $\circ$ denotes superposition,  $L_i : \mathbb{R}^{s_{i-1}} \to \mathbb{R}^{s_i}, i\in\{1,2,3,4\}$ represent parametric linear transformations and translations, known as linear layers, $\BN_i : \mathbb{R}^{s_i} \to \mathbb{R}^{s_i}, i\in\{1,2,3\}$ denote batch normalization layers,  
and $\text{ReLU}_i : \mathbb{R}^{s_i} \to \mathbb{R}^{s_i}, i\in\{1,2,3\}$ the rectified linear unit activation function. The integers $s_{i-1},s_i \in \N=\{1,2,\ldots\}$ denote the dimensionality of the input and output sizes of the layers. 
A schematic representation of the entire generator network architecture as well as the choices of $s_i \in \N$ for $i\in \{1,2,3,4\}$ are shown in Figure~\ref{fig:generator_architecture}.

For each $i\in\{1,2,3,4\}$,
the $i$-th linear layer $L_i$ performs a transformation $L_i : \mathbb{R}^{s_{i-1}} \to \mathbb{R}^{s_{i}}$ given by
\begin{align} L_i(x) = W_ix + b_i, \end{align}
for each $x\in\R^{s_{i-1}}$, where $W_i \in \mathbb{R}^{{s_{i}} \times {s_{i-1}}}$ is a trainable weight matrix, and $b_i \in \mathbb{R}^{s_i}$ is a trainable bias vector. Here, ${s_{i-1}}$ corresponds to the input size, and ${s_{i}}$ is the output size of the $i$-th linear layer $L_i$. 

Moreover,
to stabilize training and improve convergence, batch normalization layers $\BN_i : \mathbb{R}^{s_i} \to \mathbb{R}^{s_i}$ are utilized for $i\in\{1,2,3\}$. Note that batch normalization layers rescale the value of their input to match a trainable mean and trainable variance while maintaining the dimensionality $s_i$ of the input, see \cite{ioffe2015batch} for further details.

Additionally, non-parametric non-linear ReLU activation functions $\text{ReLU}_i : \mathbb{R}^{s_i} \to \mathbb{R}^{s_i}$, where $i\in\{1,2,3\}$, allow the network to represent complex, non-linear dependencies. Note that the rectified linear unit activation functions ReLU$_i$ are defined componentwise, i.e.,
\begin{align} \text{ReLU}_i(x) =\bigl( \max\{0, x_1\},\ldots,\max\{0,x_{s_i}\}\bigr)  
\end{align}
for any $i\in\{1,2,3\}$ and  $x=(x_1,\ldots,x_{s_i})\in \R^{s_i}$.
For an overview of the input and output sizes, as well as  the numbers of trainable parameters, see Table~\ref{tab:fcsh_generator_architecture}.

\begin{table}[H]
    \centering
    \tiny
    \renewcommand{\arraystretch}{1.5} 
    \begin{tabular}{|>{\centering\arraybackslash}m{3cm}|>{\centering\arraybackslash}m{3cm}|>{\centering\arraybackslash}m{4cm}|}
        \hline
        \textbf{Transformation Type} &  \textbf{Output Size $s_i$} & \textbf{Trainable Parameters} \\
        \hline
        Input    & 49 & - \\
        \hline
        $\text{ReLU}_1 \circ B_1 \circ L_1$  & 64 & $(49 \cdot 64) + 64 + 128 = 3328$ \\
        \hline
        $\text{ReLU}_2 \circ B_2 \circ L_2$  & 128 & $(64 \cdot 128) + 128 + 256 = 8576$ \\
        \hline
        $\text{ReLU}_3 \circ B_3 \circ L_3$  & 256 & $(128 \cdot 256) + 256 + 512 = 33,536$ \\
        \hline
        $L_4$  & 49 & $(256 \cdot 49) + 49 = 12,593$ \\
        \hline
        
    \end{tabular}
    \caption{Input and output sizes as well as the corresponding numbers of trainable parameters for the layers of  the neural network that generates the spherical harmonics coefficients $a_{\ell m}$ from standard normal distributed noise.}
    \label{tab:fcsh_generator_architecture}
\end{table}

\subsubsection{Loss function}\label{sec:loss} 

Recall that the parameters of the neural network $f\colon \R^{49}\to  \R^{49}$ given in Eq.~\eqref{def.neu.net} have to be optimized to ensure that the realizations of the particle hull model $Z$, as defined in Eq.~\eqref{zet.lin.com}, are statistically similar to the measured data. 
Importantly, this objective cannot be achieved solely by assessing the authenticity of individual realizations $\rthreeD\colon [0,2\pi)\times[0,\pi)\to\R$ of $\randomField$, 
as such an approach would, for instance, fail to account for the distribution of particle hull sizes.
To take this into  account, a loss function 
$l \colon \mathcal{F}_\mathrm{3D}^b \times \mathcal{F}_\mathrm{2D}^b \to \mathbb{R}$ 
is considered, where  $\mathcal{F}_\mathrm{3D}$ and $\mathcal{F}_\mathrm{2D}$ denote the sets of all radius functions $\rthreeD\colon [0,2\pi)\times[0,\pi)\to\R$ and $r_\mathrm{2D}\colon [0,2\pi)\to\R$, respectively, and $b\in \N$ is some batch parameter.
The loss function evaluates the authenticity of a sequence $S_\mathrm{3D} \in \mathcal{F}_\mathrm{3D}^b$ of $b$ realizations of the spherical harmonics model $\randomField$ by comparing them to a sequence $M_\mathrm{2D} \in \mathcal{F}_\mathrm{2D}^b$ of $b$ radius functions derived from measured data. The value $l(S_\mathrm{3D}, M_\mathrm{2D})$ of the loss function corresponding to the pair $(S_\mathrm{3D}, M_\mathrm{2D})$ is defined as
\begin{align}
    l(S_\mathrm{3D}, M_\mathrm{2D}) = 
    l_{1}(\sigma(S_\mathrm{3D}), M_\mathrm{2D}) 
    + l_{2}(\sigma(S_\mathrm{3D}), M_\mathrm{2D}) 
    + l_{3}(\sigma(S_\mathrm{3D}), M_\mathrm{2D}) 
    + l_4(\sigma(S_\mathrm{3D})) 
    +  l_{5}(S_\mathrm{3D}),
    \label{eq:loss}
\end{align}
where $\sigma(S_\mathrm{3D}) \in \mathcal{F}_\mathrm{2D}^b$ is the sequence of 2D radius functions obtained by computing the elementwise projections of the elements of $S_\mathrm{3D}$, as described in Section~\ref{sec:radial_rep}. The functions $l_{1}$, $l_{2}$, $l_{3}\colon\mathcal{F}_\mathrm{2D}^b \times \mathcal{F}_\mathrm{2D}^b \to \mathbb{R}$ appearing on the right-hand side of Eq.~\eqref{eq:loss} quantify the statistical similarity between $\sigma(S_\mathrm{3D})$ and $M_\mathrm{2D}$ based on interpretable descriptors. Furhtermore, the term $l_\mathrm{4}\colon\mathcal{F}_\mathrm{2D}^b \to \mathbb{R}$ rates the authenticity   of simulated projections by the so-called  discriminator, i.e., by non-interpretable descriptors learned by a second neural network. Finally, $l_{5}\colon \mathcal{F}_\mathrm{3D}^b \to \mathbb{R}$ penalizes artifacts in the 3D structures of $S_\mathrm{3D}$ that cannot be observed in the 2D projections. 

To measure the similarity of two sequences of radius functions, the loss functions $l_1,l_2,l_3$ utilize the concept of order statistics. Therefore, for each $j\in\{1,2,3\}$, let $d_j:\mathcal{F}_\mathrm{2D} \to \R$
be a mapping that assigns a scalar descriptor value $d_j(r)$ to each radius function $r\in \mathcal{F}_\mathrm{2D}$. For  a sequence of radius functions $R=(r_1,\ldots,r_b)\in\mathcal{F}_\mathrm{2D}^b$, we denote the elementwise evaluation of $d_j$ on the  components of $R$ simply  by
$d_j(R)=(d_j(r_1),\ldots,d_j(r_b))$ if no confusion is possible.
Furthermore, for each $k\in\{1,\ldots,b\}$, let 
$d_j(R)_{(k)}$  denote the $k$-th smallest value of $d_j(R)=(d_j(r_1),\ldots,d_j(r_b))$.   
To further simplify the notation, we will briefly write $S_\mathrm{2D}$ instead of  $\sigma(S_\mathrm{3D})$ in the following. 
The similarity of the sequences $d_j(S_\mathrm{2D})$ and $d_j(M_\mathrm{2D})$ of descriptor values is then given by their mean absolute difference, i.e.,

\begin{align}
    \label{mea.abs.dif}
    l_{j}(S_\mathrm{2D},M_\mathrm{2D}) = \frac{1}{b}\sum_{k=1}^{b}|d_j(S_\mathrm{2D})_{(k)}-d_j(M_\mathrm{2D})_{(k)}|, \qquad\mbox{for  $j \in \{1,2,3\}$,}
\end{align}
where the mappings $d_1,d_2,d_3:\mathcal{F}_\mathrm{2D} \to \R$  characterize the shape of radius functions in terms of their mean radius, circumference, and centeredness, respectively. 
More precisely, it holds  that 
\begin{align}
    d_1(r
    )&=\frac{1}{\kappa}\sum_{i=0} ^{\kappa-1} r(\phi_i),\label{eq:mean_radius}\\
    d_2(r)&=\sum_{i=0} ^{\kappa -1} r(\phi_i)^2+r(\phi_{i+1})^2 - 2r(\phi_i)r(\phi_{i+1}) \cos(\frac{2 \pi}{\kappa})\label{eq:circumference}\\
    d_3(r)&= \sum_{i=0}^{\frac{\kappa}{2}-1}| r(\phi_i)- r(\phi_{\frac{\kappa}{2}+i})|,\label{eq:centeredness}
\end{align}
for any $r\in\mathcal{F}_\mathrm{2D}$, where  $\phi_i = \frac{2\pi i}{\kappa},i\in\{0,\ldots\kappa-1\}$ are equidistant angles on which the radius functions are evaluated, for some even number $\kappa\in\{2,4,\ldots\}$.
Thus, $d_1(r)$ is the average value of the radius function $r:[0,2\pi)\to[0,\infty)$ across $\kappa$ equidistant angles, the circumference $d_2(r)$ approximates the perimeter of the radius function in $r$ using polylines formed by $\kappa$ segments between adjacent angles, and the centeredness $d_3(r)$ quantifies the alignment between the origin of the radius function $r$ and the center of mass of the area enclosed by $r$. Due to the heuristic choice of setting the origin of the radius functions derived from TEM images to center of mass of the segmented area, a significant discrepancy between these two values in the simulated radius functions could lead to results that produce similar particles, which, however, may be distinguishable from the measured ones due to their radial representation. Thus, the loss function $l_3$ in Eq.~\eqref{eq:loss} is primarily used to accelerate convergence  during model fitting, especially when considering the discriminator-based loss introduced in Eq.~\eqref{eq_discriminator_loss} below.  
As already mentioned in Section~\ref{sec:radial_rep},
the value of $\kappa$ is set to 16 in order to achieve high computational speed.

However, there are many more 2D descriptors, most of which are significantly more complex than the mappings $d_1, d_2, d_3$ considered in Eqs.~\eqref{eq:mean_radius}--\eqref{eq:centeredness}.  To ensure high statistical similarity of the simulated and measured data, the distributions of these more complex descriptors must also match between the simulated and measured radius functions. To account for such additional descriptors, the loss function $l_4\colon \mathcal{F}_\mathrm{2D}^b \to \mathbb{R}$ is considered in Eq.~\eqref{eq:loss}, which is defined as a high parametric function, whose parameters are continuously updated to compute non-interpretable descriptors that reveal discrepancies between simulated and measured radius functions.
More specifically, the loss function $l_4$ is based on a second neural network, the so-called discriminator $D\colon \R^{\kappa} \to \mathbb{R}$, where
\begin{align}
    l_4(R) = \frac{1}{b}\sum_{r \in R} D(r(\phi_0), \ldots, r(\phi_{\kappa-1} )),\label{eq_discriminator_loss}
\end{align}
for each sequence $R\in\mathcal{F}_\mathrm{2D}^b$  of   radius functions, with $\phi_i = \frac{2\pi i}{\kappa}, i \in \{0, \ldots, \kappa-1\}$ specifying the discretization of the radius functions. 
The neural network $D$ is trained to distinguish between simulated and measured radius functions, thereby implicitly learning non-interpretable descriptors whose values differ between simulated and measured radius functions. More precisely, the network $D$ is trained to assign high output values to radius functions derived from simulated data and low output values to those from measured data. Consequently, minimizing the output of $D$ for simulated radii functions is desirable for fitting the particle hull model.

Note that the radius functions  in Eq.~\eqref{eq_discriminator_loss} are  periodic with period $2\pi$. This necessitates a specific choice of $D$ that accounts for this periodicity. Thus, $D$ is chosen as a discriminator network that employs 1D convolutional layers, i.e., layers designed for local feature extraction, able to handle input with periodic features.

Like the generator network $f$ given in Eq.~\eqref{def.neu.net}, the discriminator network $D$ is composed of simple parametric functions, referred to as layers. More precisely, the discriminator network $D\colon \R^{\kappa}\to \R$ can be written down as
\begin{align}\label{def.dis.net}
    D &= \mathrm{MEAN}_{19}\circ \mathrm{Conv}_{18}\circ \mathrm{CB}_{15}\circ \mathrm{MaxPool}_{14}\circ \mathrm{CB}_{11}\circ \mathrm{CB}_8\circ\mathrm{MaxPool}_7\circ \mathrm{CB}_4\circ \mathrm{CB}_1,
    \end{align}
    where
    \begin{align}
    \mathrm{CB}_i&=\mathrm{ReLU}_{i+2}\circ\BN_{i+1}\circ\mathrm{Conv}_i,
\end{align}
and $\mathrm{Conv}_i\colon\R^{c_{i-1}}\to\R^{c_{i}},i\in\{1,4,8,11,15,18\}$ are convolutional layers with a kernel size of 3 and periodic padding for extracting local and periodic features,
$\mathrm{BN}_i\colon\R^{c_{i-1}}\to\R^{c_{i}},i\in\{2,5,9,12,16\}$ are batch normalization layers used to stabilize training and accelerate convergence, 
$\mathrm{ReLU}_i\colon\R^{c_{i-1}}\to\R^{c_{i}},i\in\{3,6,10,13,17\}$ are ReLU activation functions used to introduce non-linearity into the model,  $\mathrm{MaxPool}_i\colon\R^{c_{i-1}}\to\R^{ c_{i}},i\in \{7,14\}$ are max pooling layers with a stride of 2 for dimensionality reduction, and $\mathrm{MEAN}_{19}\colon\R^{c_{18}}\to\R^{ c_{19}}$ computes the arithmetic mean in order to return a scalar network prediction per radius function. 

\begin{figure}[H]
    \centering
    \includegraphics[width=.39\linewidth]{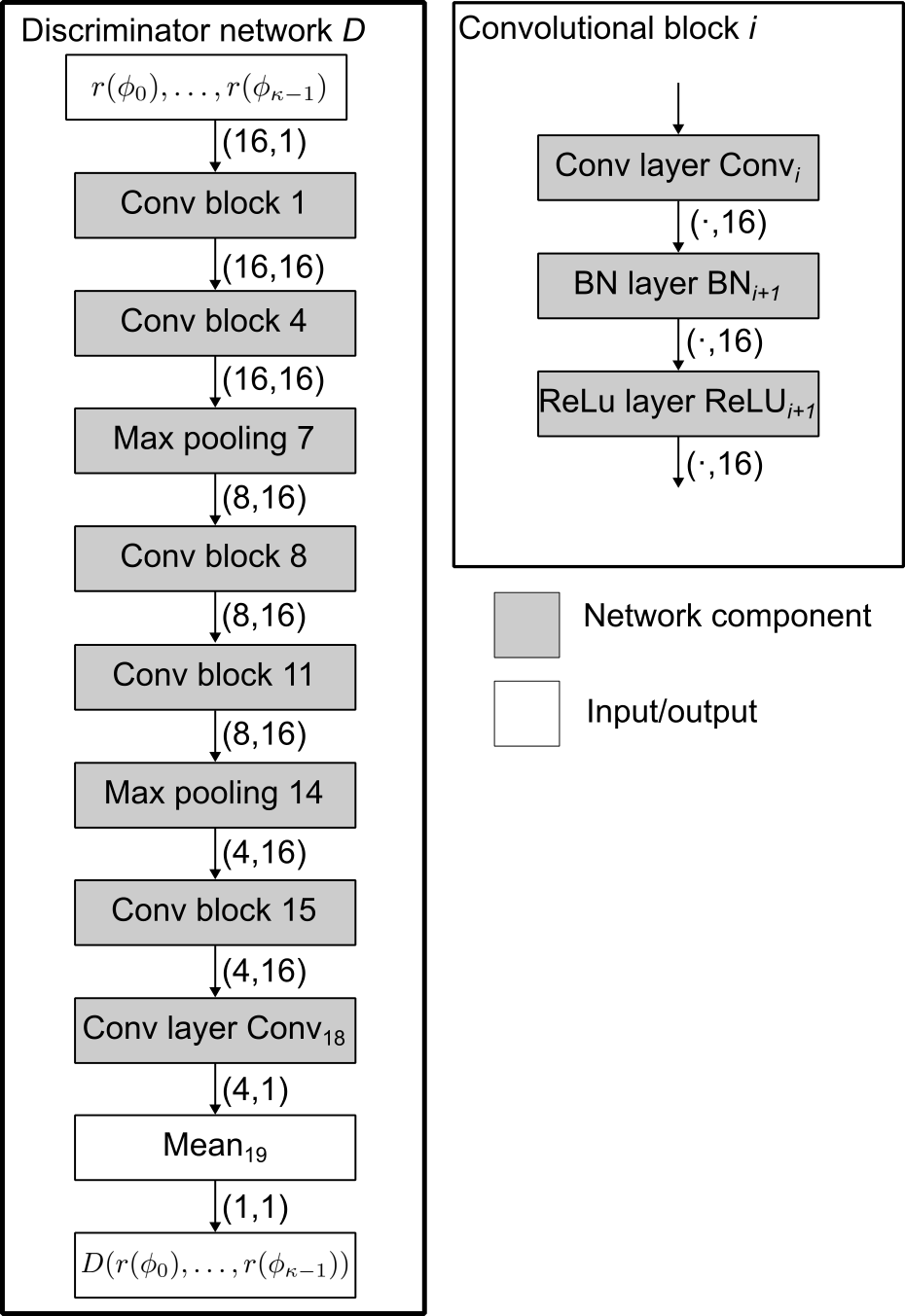}
   
    \caption{Architecture of the discriminator network $D$,    designed to distinguish between simulated and measured radius functions. The numbers correspond to the input and output dimensions of the respective layers.
    }
    \label{tab:discriminator_architecture}
\end{figure}

However, unlike the generator network $f$ given in Eq.~\eqref{def.neu.net}, each layer of the discriminator network $D$ has a pair of input and output dimensionalities $(c_{i-1}, c_i) \in \mathbb{N} \times \mathbb{N}$, given by two positive integers which specify the number of rows and columns of the respective matrix. The first integer $c_{i-1}$ corresponds to the resolution of the radius function, determining how finely the radius is discretized, while the second integer $c_i$ represents the number of channels computed per angle, capturing different learned characteristics. The input $(r(\phi_0), \ldots, r(\phi_{\kappa-1} ))\in \R^\kappa$ of $D$ has a corresponding dimensionality of $(\kappa,1)$ as each angle has only one corresponding channel, containing the value of the radius function. As the layers are applied to the input, the number of  angles decreases, enabling the extraction of more global features, while the number of channels increases to 16, allowing for the computation of more complex features. This is analogous to the decreasing resolution and increasing number of channels in a 2D CNN feature map in computer vision~\cite{Goodfellow-et-al-2016}.
An overview of the values $c_i, i \in \{0,\ldots,19\}$ of these dimensions and an illustration of the network architecture are provided in Figure~\ref{tab:discriminator_architecture}.

Finally, the loss function  $l_{5}\colon \mathcal{F}_\mathrm{3D}^b \to \mathbb{R}$ considered in Eq.~\eqref{eq:loss}  is formally defined as follows. Note that very fine  surface features of a simulated 3D particle hull  are neither of interest for the particle hull model, since these are subsequently modeled by a sphere packing model, see Section~\ref{sec:sphere_packing_model}, nor these features are observable in the low-resolution training data.
Thus, the loss function $l_{5}\colon \mathcal{F}_\mathrm{3D}^b \to \mathbb{R}$ is introduced to avoid a very high surface roughness, which would lead to a decrease in sphere packing speed. In contrast to the previously introduced loss terms,  the loss function $l_{5}$ does not evaluate virtually projected radius functions, but non-projected radius functions. More precisely, for a sequence $R\in \mathcal{F}_\mathrm{3D}^b$ of radius functions  $r\colon [0,2\pi)\times [0, \pi)\to \R$ of 3D particle hulls, the roughness loss $l_\mathrm{5}(R)$ is given by
\begin{align}
    l_5(R) = \frac{1}{10b}\sum_{r \in R} \sum_{i =0}^{\kappa-1} \sum_{j=0}^7 \max \left\{0, \left|\frac{r(\phi_i,\theta_{j-1})+r(\phi_i,\theta_{j+1})}{2} - r(\phi_i,\theta_j) \right|-2\right\}, \label{rou.los.ter}
\end{align}
where $\phi_i = \frac{2 \pi i}{\kappa}, i\in \{0,\ldots,\kappa-1\},~ \theta_j=\frac{\pi j}{8},j\in\{0,\ldots,7\}$, and $\frac{1}{10b}$ is some scaling constant in order to adjust the magnitude of the loss.
Intuitively speaking, 
the roughness loss $l_\mathrm{5}(R)$ 
given in Eq.~\eqref{rou.los.ter} penalizes deviations of radius function  $r\colon [0,2\pi)\times [0, \pi)\to \R$ at position $(\phi_i,\theta_j)$, if they lead to values $r(\phi_i,\theta_j)$ that deviate by more than 2 units from the average value of the radius at the neighboring positions $(\phi_i,\theta_{j-1})$ and $(\phi_i,\theta_{j+1})$.\\

\subsubsection{Fitting procedure}\label{sec:net_training}
In order to enable the generation of realistic 3D particles with the particle hull model $Z$ introduced in Section~\ref{sec:outer_hull_model}, the
parameters of the  spherical harmonics coefficients generating network $f$ given in Eq.~\eqref{def.neu.net}
have to be fitted. This is done by minimizing the sum on the right-hand side of Eq.~\eqref{eq:loss}, where a gradient descent method is used,  utilizing an Adam optimizer~\cite{kingma2014adam} with a learning rate of $0.01$. The number $b\in\N$, introduced in Section~\ref{sec:loss}, of simulated particle hulls per gradient descent step  and thus the sizes of $S_\mathrm{3D},S_\mathrm{2D}$ and $M_\mathrm{2D}$ is set to $512$. Note that the value of $b$ has to be chosen sufficiently large such that $S_\mathrm{2D}$ and $M_\mathrm{2D}$ are representative and, thus, the losses $l_1,l_2,l_3$ based on order statistics, see  Eq.~\eqref{mea.abs.dif},
are meaningful. While training the generator  network $f$, also a training of the discriminator network $D$ given in Eq.~\eqref{def.dis.net}  
takes place, which  results in a continuous refinement of the non-interpretable descriptors defined by $D$. The training of $D$ is done by minimizing the loss function $|l_4(M_\mathrm{2D})-l_4(S_\mathrm{2D})|$, using a gradient descent method and an Adam optimizer with a learning rate of $0.0001$. The steps of this minimization are done at every 10-th step of the generator training with a batch size of $b=256$, i.e., each training step of $D$ is based on 2560 simulated and 2560 measured radius functions.

\begin{figure}[ht!]
    \centering
    \includegraphics[width =.9\textwidth]{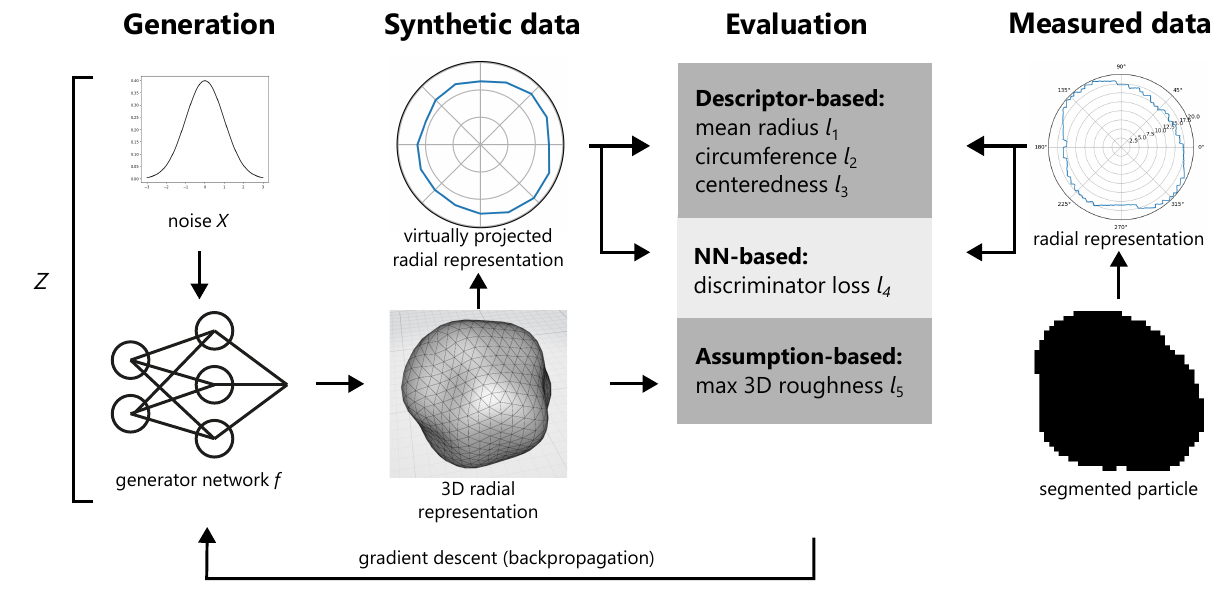}
    \caption{Visualization of the fitting procedure for the particle hull model $Z$. Recall that the radius functions $\rtwoD$, given in Eq.~\eqref{eq:projection},  of the projections of synthetic data are represented by means of $\kappa = 16$ angles.}
    \label{fig:approach_hull}
\end{figure}

Note that the parameters of the generator network $f$ should be optimized such that the particle hull model $\randomField$ is isotropic. Furthermore, the discriminator should not be able to use any anisotropic features present in the measured data (due to their pixel based representation). 
For these reasons, the radius functions $r_\mathrm{2D}$ derived from measured data, as well as the simulated radius functions $\rthreeD$, are rotated uniformly (in 2D and 3D, respectively) before being discretized and used for training. An overview of the whole fitting procedure for $\randomField$ is given in Figure~\ref{fig:approach_hull}.

\subsection{Sterological fitting of the sphere packing model}\label{sec:sphere_packing_fitting}

To achieve realistic surfaces of simulated 3D particles using the sphere packing model described in Section~\ref{sec:sphere_packing_model}, the number $n$ of spheres placed within a particle hull, their size $\radiusSpherePacing$, as well as the magnitude $\delta$ of the applied morphological closing have to be determined such that the fine surface features match those observed in measured image data on a fine length scale, see Figure~\ref{fig:high_resolved}.

To estimate the radius $\radiusSpherePacing$ that adequately represents the outline features observed in Figure~\ref{fig:high_resolved}c, the representation of the  outline of  projected 3D particles is first changed. Specifically, the sequence $O=(o_1,\ldots,o_k) \subset \Z^2$ of consecutive pixel positions of the outline pixels is constructed. In other words, for each $i\in \{1,\ldots,k\}$, the pixel $o_i$
belongs to the particle and has at least one neighboring pixel that does not belong to the particle. Furthermore, the consecutive pixel positions $o_i$ and $o_{i+1}$ for $i\in \{1,\ldots,l-1\}$ have a distance of 1 with respect to the maximum norm. 
As a second step, each pixel position $o_i\in O$ is labeled as either being part of a convex curvature or a concave curvature. Specifically,  for each $i\in \{1,\ldots,k\}$, the outline pixel $o_i$ is labeled based on the cross product $o_i^+ \times o_i^-$ of vectors $o_i^-$ and $o_i^+$, given by
\begin{equation}
    o_i^- = o_i-o_{i-10}\qquad\mbox{and}\qquad
    o_i^+ = o_{i+10}-o_i,
\end{equation}
where we set $o_i=o_1$ for $i<1$ and $o_i=o_k$ for $i>k$. Note that instead of considering the direct predecessor of $o_i$, the nearest 9 pixel positions are skipped. This is done in order to be more resistant to noise occurring due to the pixel based representation of the outline.
The label $l(o_i)$ of $o_i$ is then given by $l(o_i)=\sign(o_i^+ \times o_i^-)$, where $\sign\colon \R \to \{-1,0,1\}$ denotes the sign function. A label of 1 corresponds to a convex curvature, and a label of $-1$ to a concave curvature.

To estimate the radius $\radiusSpherePacing$, spheres are fitted to the inclusion-maximum sequences of convex outline pixel positions in $O$. For any $l,m>0$, a subsequence $I=(o_l, o_{l+1}, \ldots, o_{l+m})\subset O$  is called a an inclusion-maximum sequence if  
\begin{equation}
    l(o_i)=1 \text{ for all }  i \in \{l, \ldots, l+m\} \qquad\mbox{and}\qquad
    l(o_{l-1})\neq1, l(o_{l+m+1})\neq1,
\end{equation}
where $o_0$ and $o_{k+1}$ are considered to not belong to a convex curvature, i.e., $l(o_0) = l(o_{k+1}) \neq 1$.
For each inclusion-maximum sequence $I=(o_l,\ldots, o_{l+m})$, the radius $r_I>0$ of a sphere is computed  that represents this sequence in the best possible way.  
More precisely, a pair $(c_I,r_I)$ with $c_I\in\R^2$ and $r_I>0$ is determined, which is given by
\begin{align}
    (c_I,r_I) = \text{argmin}_{(c,r)\in\R^2\times[0,\infty)}\sum_{i=l}^{l+m} ( |o_i-c| - r )^2,
\end{align}
where  $(c,r)$ denotes the center $c\in \R^2$ of a circle and its radius $r>0$, and $|x|=\sqrt{x_1^2+x_2^2}$ is the Euclidean norm of $x=(x_1,x_2)\in\R^2$. 
The sphere radius $\radiusSpherePacing>0$ is then  given by 
\begin{align}
    \radiusSpherePacing = \frac{\sum_{I \in \mathcal{C}} r_I\,\# I}{\sum_{I \in \mathcal{C}} \# I}, 
\end{align}
where $\mathcal{C}$ denotes the family of all inclusion-maximum sequences  $I=(o_l, o_{l+1}, \ldots, o_{l+m})\subset O$ of convex outline pixel positions in $O$, and $\# I =m$ is the length of the sequence $I\in\mathcal{C}$.

The number $n$ of placed spheres heavily influences the volume of the resulting particle. Thus, the volume of the realization of the outer hull model is used as a basis for the estimation of $n$. More precisely, $n$ is set as the number of spheres that have a total of volume which approximates the volume of the outer hull realization, i.e., $n$ is set to the closest integer to $\frac{3V(\widehat{r}_{\mathrm{3D}})}{4\pi \radiusSpherePacing^3}$,
where $\widehat{r}_{\mathrm{3D}}$ is the radius function of the underlying particle hull, and $V(\widehat{r}_{\mathrm{3D}})$ denotes  its enclosed volume.

\begin{figure}[ht]
\centering 
     \begin{subfigure}{0.35\textwidth}
     \centering
         \includegraphics[width = \textwidth]{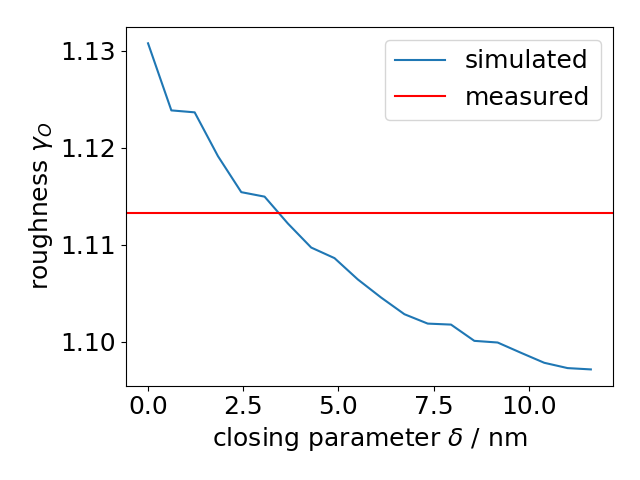}
     \end{subfigure}
     \hfill

    \caption{ Interplay between the magnitude of morphological closing $\delta$ and the roughness $\gamma_O$.  
The blue curve represents the roughness of the outline of a simulated particle as a function of the magnitude $\delta$ of morphological closing  applied to a given sphere packing. The red line shows the roughness obtained from high-resolution image data.}
    \label{fig:closing_roughness}
\end{figure}

The third  parameter of the sphere packing model, i.e., the magnitude $\delta$ of the morphological closing, is estimated by considering the so-called roughness of the  outline, where 
the roughness quantifies the depth of indents within the particle outline by comparing the length of the traced particle outline between two points to the length of their straight-line distance.
More precisely, for 
the sequence $O=(o_1,\ldots,o_k) \subset \Z^2$ of  outline pixels, the roughness $\roughness_O$ is defined by\begin{equation}
    \roughness_O = \frac{1}{k}\sum_{i=1}^{k-1} \frac{\sum_{j=i}^{\nu(i)-1}|o_{j}-o_{j+1}|}{|o_i-o_{\nu(i)}|}\qquad\mbox{with}\qquad  \nu(i) =\min \{j \in \{i+1,\ldots,k\} \colon |o_i-o_{j}| \geq h \text{ or } j=k \},\label{ew:roughness}
\end{equation}
where $h>0$ is some resolution parameter which is set to $h=\radiusSpherePacing$.
The procedure for determining the closing parameter $\delta$ is as follows. First, $\delta_-=0$ nm is set to the smallest possible value of $\delta$, and $\delta_+ = 20$ nm to the highest expected value of $\delta$.  
Then, a 3D particle is generated with $\delta = \frac{\delta_- + \delta_+}{2}$, for which
a virtual pixel-based projection is  computed,  using the same pixel size as in the high-resolution TEM images. The roughness $ \roughness_O$ of this projection, given in Eq.~\eqref{ew:roughness}, is compared to that of particles observed in TEM images. If the roughness of the simulated particle outline exceeds that of  measured particles, $\delta_-$ is updated to $\delta$, otherwise $\delta_+$ is updated to $\delta$.  
This procedure always terminates because the roughness $ \roughness_O$ of the simulated particle outline decreases monotonously with increasing $\delta$ (see Figure~\ref{fig:closing_roughness}) and can only take a finite number of values due to the pixel-based representation of $O\subset\Z^2$.

\section{Results and Discussion}\label{sec:results}
\noindent
In this section, the projections of realizations of the stochastic models introduced in Section~\ref{sec.two.sta}, i.e., the pre-stage particle hull model and the sphere packing model, are statistically compared with measured (projections of) particle outlines. To do so, the radius functions of $1\,000$~realizations of the fitted particle  hull model and the fitted sphere packing model, as well as the radius functions of particle outlines observed in the low-resolution image data, see Figure~\ref{fig:preprocessing}, are computed. For computational purposes, like in Section~\ref{sec:stereological_hull_fitting}, these radius functions are evaluated on a finite number of equidistant angles. However, in the present section, we do not consider 16,  but 128 equidistant angles.
The radius functions discretized in this way  are then compared in terms of their circumference $d_2$ given in Eq.~\eqref{eq:circumference}. Furthermore, the following descriptors of radius functions are considered: the minimum diameter  $\minDiameter$, the maximum diameter $\maxDiameter$, the  elongation $\elongation$ and the area $\area$, where
for the 128 equidistant evaluation points $\phi_i=\frac{2i\pi}{\kappa}, i\in \{0,\ldots,\kappa\}$, with $\kappa = 128$ and a radius function $r\colon [0,2\pi)\to [0,\infty)$  it holds that
\begin{align}
     \minDiameter(r)&= \min \bigl \{  r(\phi_i)+r(\phi_{i+ 64}) \colon i  \in \{0, \ldots,  63\} \bigr\} \label{eq:minDiameter}\\
     \maxDiameter(r)&= \max \bigl \{  r(\phi_i)+r(\phi_{i+ 64}) \colon i  \in \{0, \ldots,  63\} \bigr\}\\
     \elongation(r)&=\frac{\minDiameter(r)}{\maxDiameter(r)}\\
     \area(r)      &= \sum_{i=0} ^{127} \frac{r(\phi_i)r(\phi_{i+1}) \sin(\frac{2\pi}{128}) }{2},\label{eq:elongation}
\end{align}
wih $\phi_{128}=\phi_0$.

Note that the circumference $d_2$ given in Eq.~\eqref{eq:circumference}  was already used to fit the particle hull model, but on a smaller number of evaluation points. However, the other four descriptors given in Eqs.~\eqref{eq:minDiameter}-\eqref{eq:elongation} do not correspond to any of the loss functions $l_1,l_2,l_3$ considered in Eq.~\eqref{mea.abs.dif}. Intuitively speaking, the elongation $\elongation(r)$ of a radius function $r$ is the ratio of the lengths of the shortest to the longest line passing through the center of $r$, while $\area(r)$ is an approximation of the area enclosed by $r,$ computed as the sum of the areas of 128 triangles. Each triangle has sides of length $r(\phi_i)$ and $r(\phi_{i+1})$, and encloses an angle of $\frac{2\pi}{128}$.

\begin{figure}[H]
    \centering
    \includegraphics[width=\textwidth]{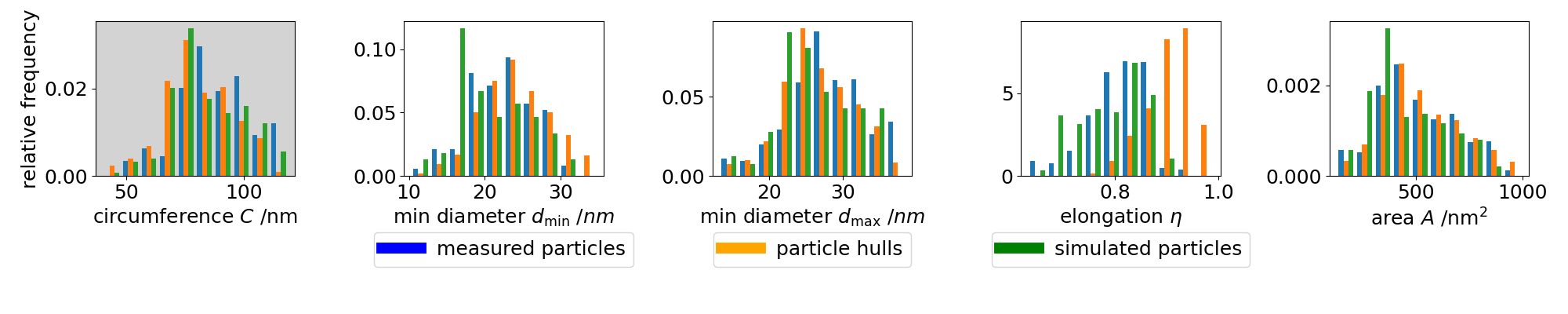}
    \caption{Histograms of  descriptors of projected outlines for measured particles  (blue), as well as for realizations drawn from the fitted particle hull model (orange) and  the fitted sphere packing model (green). Recall that the circumference (gray background) is used to calibrate the particle hull model, whereas the other four descriptors are not directly used for model calibration.}
    \label{fig:results}
\end{figure}

As already mentioned above, the number of evaluation points considered in this section is significantly higher than the number of angles used in Section~\ref{sec:stereological_hull_fitting} to fit the particle hull model  ($128 \gg 16$). However, despite the much higher resolution of evaluation points, the particle hull model still achieves a quite good fit,  see the blue and orange bars in Figure~\ref{fig:results}. In particular, the particle hull model effectively captures the circumference $d_2$, the minimum and maximum diameters, $d_\mathrm{min}$ and $d_\mathrm{max}$, and the area $\area$ of projected particle outlines. On the other hand, it tends to generate particle hulls with projections that are more circular compared to the measured ones.
The particles generated by the sphere packing model described in Section~\ref{sec:sphere_packing_model} demonstrate a similarly good quality of fit with respect to the descriptors $d_2$, $d_\mathrm{min}$, $d_\mathrm{max}$, and $\area$. Additionally, due to the modeling of the particle surface using packed spheres and morphological closing, more realistic indents are produced. This results in a better representation of the correlation between $d_\mathrm{min}$ and $d_\mathrm{max}$ in terms of their ratio, i.e.,  elongation $\elongation$, as shown in Figure~\ref{fig:results}.

The high similarity of  descriptor distributions for simulated and measured particle outlines shown in Figure~\ref{fig:results} 
gives rise to the conjecture that the fitted sphere packing model reflects not only the  descriptor distributions of projected 2D outlines, but also those of the 3D particles themselves. 
However, the extent to which this model accurately represents the 3D morphology of catalytic particles remains an open question, as measuring a statistically representative number of these particles at sufficiently high resolution in 3D is not feasible. In our future work, we will evaluate this at least on a small number of particles measured by means of high-resolution 3D imaging.

\begin{figure}[ht!]
    \centering
    \includegraphics[width=\linewidth]{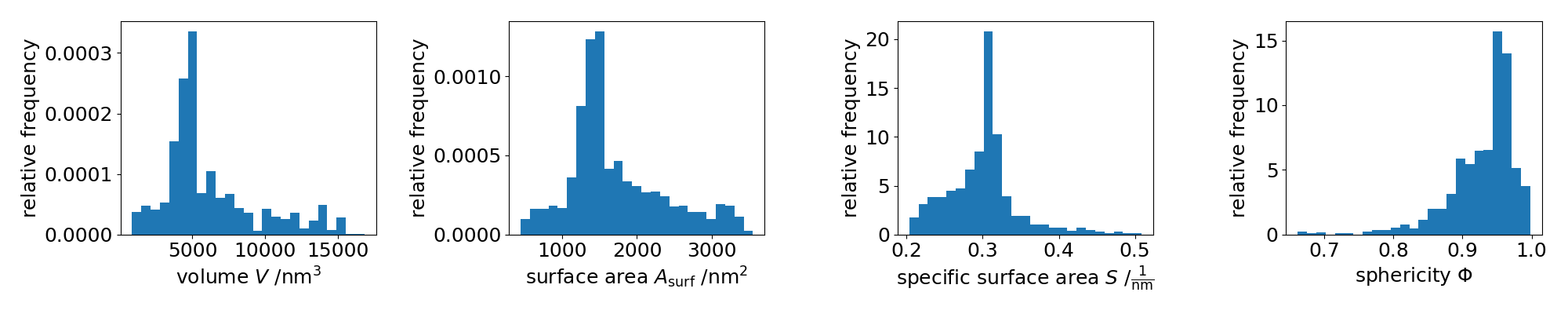}
    \caption{Histograms  of the  volume $\volume$, the surface area $\surfaceArea$, the specific surface area $\specificSurface$, and the sphericity $\sphericity$ of simulated 3D particles, drawn from the sphere packing model.    
   }
    \label{fig:3D_descriptor_prediction}
\end{figure}

Nevertheless, in the following, we consider some descriptors of simulated 3D particles, relevant in catalysis, by drawing realizations from the sphere packing model. More precisely, $1\, 000$ realizations of the sphere packing model are used to investigate the distributions of the  volume $\volume$, the surface area $\surfaceArea$, the specific surface area $\specificSurface$, and the sphericity $\sphericity$ of 3D particles. The values of these descriptors are computed on a voxel-based particle representation with a voxel side length of 0.102 nm, corresponding to the resolution displayed in Figure~\ref{fig:high_resolved}. Note that the volume $\volume$ is determined by simply counting the number of voxels belonging to a particle, whereas the surface area $\surfaceArea$ is determined by an approach presented in~\cite{schladitz.2007}. The specific surface area $\specificSurface$ and the  sphericity $\sphericity$ are then given by well-known formulas, i.e., 
\begin{equation}
\specificSurface=\frac{\surfaceArea}{\volume} \qquad\mbox{and}\qquad \sphericity=\frac{\pi^\frac13(6\volume)^\frac{2}{3}}{\surfaceArea}.
\end{equation}
The resulting histograms of $\volume$, 
$\surfaceArea$, 
$\specificSurface$ and $\sphericity$
are shown in Figure~\ref{fig:3D_descriptor_prediction}. They  are unimodal and non-symmetric, which is a typical phenomenon for this kind of particle descriptors~\cite{furat2021artificial,furat2024multidimensional}. Notably, they differ significantly from those of equally sized spheres, which are often used as a simplifying assumption for particle shape. 
This suggests that incorporating the information obtained from 2D TEM images and using the sphere packing model fitted to these data could significantly improve the prediction of catalytic performance.

\section{Conclusion and outlook}\label{sec.tio.fou}
In the present paper, a stereological modeling procedure is proposed to generate  realistic 3D catalyst particles with surface  features in the nanoscale regime. Our approach relies on the combination of two different models: a coarse but statistically representative model, calibrated by means of hundreds of particles observed in (low-resolution) TEM images with a resolution of 0,678 nm, and a second model that refines the surface features, calibrated using a few (high-resolution) TEM images with a resolution of 0.102 nm.
The presented approach enables the prediction of 3D particle descriptor distributions from 2D image data, providing a comprehensive understanding of the 3D particle morphology.

In our future work, we will use the simulated 3D particles to correlate morphological particle descriptors with  catalytic properties of the particles. For this purpose, the number of \ce{Co3O4} atoms in the simulated particles and their spatial  arrangement will be investigated, where the probability will be predicted which catalytic active sites are available and how this is represented in the catalytic performance. 
Furthermore, the low-parametric nature of the sphere packing model enables   virtual materials testing through structural scenario analyses~\cite{weber2024investigating,neumann2020quantifying}. More specifically, by systematically varying the parameters of the sphere packing model, realistic but not (yet) synthesized particles with differing surface morphologies can be simulated. Then, by predicting the catalytic performance of these 3D particles while simultaneously quantifying their surface morphology, e.g.,  by parametric modeling of multivariate particle  descriptor vectors, 
so-called structure-property relationships  can be established. This will be subject to a forthcoming research paper.

Another particularly intriguing research topic is the investigation of core-shell catalysts, where a defined pore structure can be created by a shell~\cite{strass2021bifunctional} surrounding the particle. These catalysts possess additional catalytically relevant properties, and their morphologies are reasonably well tunable due to their bottom-up synthesis approach and heterogeneous material composition. In~\cite{strass2021bifunctional}, the connection between some mean diameter-based descriptors, derived from 2D image data, and the catalytic properties of core-shell catalysts has already been investigated. It would be of interest whether 3D particle descriptors obtained by the stereological model show a stronger correlation with catalytic properties, potentially enabling more accurate predictions of the catalytic performance of core-shell particles.

\section{Acknowledgement}
\noindent

The authors gratefully acknowledge funding by the German Research Foundation (DFG) within SPP 2364 under
grant 504580586.
Furthermore, this research was   supported by DFG in the framework of project GU~1294/5-1 (Kerstin Wein, Robert Güttel).
The authors want to thank Dr. Johannes Biskupek (Institute of Material Science Electron Microscopy, Ulm University) for the preparation of the high-resolution TEM images used in this study.

\bibliographystyle{unsrt}
\bibliography{refs_Stochstics,refs_ChemicalEngineering}
\end{document}